\documentclass[12pt]{article}

\usepackage{amstext}
\usepackage{latexsym}
\usepackage{amssymb}
\usepackage{tensor} 

\def\R{\mathbb R}

\def\N{\mathbb N}

\newcommand{\eps}{\varepsilon}

\begin{document}

\newtheorem{theorem}{Theorem}[section]
\renewcommand{\thetheorem}{\arabic{section}.\arabic{theorem}}
\newtheorem{definition}[theorem]{Definition}
\newtheorem{deflem}[theorem]{Definition and Lemma}
\newtheorem{lemma}[theorem]{Lemma}
\newtheorem{example}[theorem]{Example}
\newtheorem{remark}[theorem]{Remark}
\newtheorem{remarks}[theorem]{Remarks}
\newtheorem{cor}[theorem]{Corollary}
\newtheorem{pro}[theorem]{Proposition}
\newtheorem{proposition}[theorem]{Proposition}

\renewcommand{\theequation}{\thesection.\arabic{equation}}


\title{Static solutions to the spherically symmetric
\\ Einstein-Vlasov system: \\ a particle-number-Casimir approach}

\author{H{\aa}kan Andr\'{e}asson\\
        Mathematical Sciences\\
        Chalmers University of Technology\\
        University of Gothenburg\\
        SE-41296 Gothenburg, Sweden\\
        email: hand@chalmers.se\\
        \ \\
        Markus Kunze\\
        Mathematisches Institut\\
        Universit\"at K\"oln\\
        Weyertal 86-90\\
        D-50931 K\"oln, Germany\\
        email: mkunze@math.uni-koeln.de}

\maketitle



\begin{abstract} 
Existence of spherically symmetric solutions to the Einstein-Vlasov system is well-known. 
However, it is an open problem whether or not static solutions arise as minimizers of a variational problem. 
Apart from being of interest in its own right, it is the connection to non-linear stability that gives this 
topic its importance. This problem was considered in \cite{Wol}, but as has been pointed out in \cite{AK}, 
the paper \cite{Wol} contained serious flaws. In this work we construct static solutions by solving 
the Euler-Lagrange equation for the energy density $\rho$ as a fixed point problem. The Euler-Lagrange equation 
originates from the particle number-Casimir functional introduced in \cite{Wol}. We then define a density function 
$f$ on phase space which induces the energy density 
$\rho$ and we show that it constitutes a static solution of the Einstein-Vlasov system. 
Hence we settle rigorously parts of what the author of \cite{Wol} attempted to prove.

\end{abstract}


\section{Introduction}
\setcounter{equation}{0}

The first proof of existence of spherically symmetric static solutions
to the Einstein-Vlasov system was given by Rein and Rendall in 1993, see \cite{RR}.
Several simplifications and generalizations have since then been obtained,
and we refer to \cite{A2} for a review. By now, existence of a wide class of static solutions
has been established, including a proof in the massless case \cite{AFT},
which requires very different techniques. The question whether or not static solutions
arise as minimizers of a variational problem has, on the other hand, remained open.

The aim of the work \cite{Wol} was to settle this question. However, it was shown in \cite{AK} 
that there are serious errors in \cite{Wol},
which left the problem unsolved.
In mathematical terms, the issue is if there are static solutions
to the spherically symmetric Einstein-Vlasov system that are minimizers
to the particle number-Casimir functional
\begin{equation}\label{pnC}
   {\cal D}(f)=\int_{\R^3}\int_{\R^3} e^{\lambda_f}\,(\hat{\Phi}(f)-\alpha f)\,\, dx\,dv
   =:\int_{\R^3} e^{\lambda_f}\,H(f)\,dx,
\end{equation}
as introduced by Wolansky in \cite{Wol}.
Here $f$ is the density function on phase space, $\lambda_f$ is a metric function that depends on $f$, 
and $\hat{\Phi}$ is the Legendre transform of a given ansatz function $\Phi$; lastly, $\alpha>0$ is a constant.
We have denoted by $H$ the part of the functional ${\cal D}$ for which the integration is taken over momentum space.
For a motivation and background concerning the form of the functional ${\cal D}$
we refer to \cite{Wol}. The first goal is to find a minimizer of ${\cal D}$ under a mass constraint, 
i.e., it is required that $m(r)\leq M$, for some $M>0$, where
\[
m(r):=4\pi\int_0^r s^2\rho(s) ds,
\]
and where $\rho$ is given in terms of $f$, see (\ref{rho2}) below. 
The second goal is to show that the minimizer is a static solution to the Einstein-Vlasov system. 
The route proposed in \cite{Wol} is in fact somewhat different and another variational problem is considered. 
Indeed, in this paper a certain Lagrangian $L=L(r,q,p)$ is defined and the following functional is introduced: 
\begin{equation}\label{holte}
   {\cal L}(m)=\int_0^{\infty}\, L(r,m(r),m'(r))\,dr.
\end{equation}
The variational problem is then to find a minimizer $m$ of ${\cal L}$ under the condition that $m(r)\leq M$. 
This problem is closely related to the former, as is shown in Section 6 below. If a minimizer $m$ could be obtained, 
it would then be necessary to prove that it satisfies the corresponding Euler-Lagrange equation. 
However, for doing this a major difficulty arises. It is related to the fact that the functional 
is not bounded from below: $|{\cal L}|$ becomes arbitrary large for configurations 
that are on the verge to admitting trapped surfaces. Therefore a further condition on the functions is needed 
to avoid such configurations in order to obtain a lower bound. Such a condition, however, 
drastically complicates the optimization problem, since it introduces additional ``boundaries'', 
resulting in the fact that the minimization problem is turned into an obstacle problem. 
This is one of the reasons for the gaps in \cite{Wol}, where an additional ``barrier condition''
on the mass function had been added to the set.

The Euler-Lagrange equation mentioned above can be formulated as a fixed point equation for the energy density $\rho$. 
The question we are going to address in this work is if there exists a solution $\rho_\ast$ 
to the Euler-Lagrange equation, and if it then is possible to define a density function $f_\ast$ 
that induces the energy density $\rho_\ast$, and that constitutes a static solution of the Einstein-Vlasov system. 
We give an affirmative answer to this question. 
Our result rigorously settles what the author of \cite{Wol} attempted to prove, 
and it provides a connection between a static solution and the variational problem for the density function. 
In particular, $f_\ast$ is a minimizer of the functional $H$ as introduced above, 
cf.~Section \ref{pot1} for a precise meaning. For a complete understanding, it nevertheless remains 
to show that (under reasonable constraints) there exists a minimizer 
to the full particle number-Casimir functional ${\cal D}$, which constitutes a static solution of the Einstein-Vlasov system.

Apart from being of interest in its own right, it is the connection to non-linear stability that gives this topic its importance.
In the case of the Vlasov-Poisson system (that is the Newtonian analogue of the Einstein-Vlasov system),
it is well-known that a large class of static solutions (steady states) can be obtained as minimizers
of an energy-Casimir functional, cf.~\cite{Rn2} and the references therein.
This fact has been central for proving non-linear stability of steady states of the system,
in the approach taken by Guo and Rein in 1999, see \cite{GR}.
The authors considered spherically symmetric steady states,
and the admissible class of perturbations also consisted of spherically symmetric functions.
Since then, this result has been improved, and so far the most general conditions
could be treated by Lemou, M\'{e}hats and Rapha\"{e}l in 2012, cf.~\cite{LMR},
where the class of perturbations is general and not restricted to spherical symmetry.
For a review of this topic we refer to \cite{Mou,Rn2}.

In contrast to the Newtonian case, the non-linear stability problem
for static solutions - or steady states - of the spherically symmetric Einstein-Vlasov system is open.
There are several reasons why the relativistic problem is considerably harder.
In the Newtonian case it holds that any steady state for which the density function is non-increasing
with respect to the particle energy is stable. For relativistic steady states,
under the same condition there is numerical evidence that these can be both stable or unstable,
cf.~\cite{AR} and \cite{GRS}. The main quantities that determine the stability properties
in the relativistic case seem to be the central redshift and the binding energy.

Another reason why this case is much harder is the lack of a general global existence result.
Solutions to the Vlasov-Poisson system are known to be global in time,
independent of the size of the initial data; cf.~\cite{P,LP}.
For Einstein-Vlasov the situation is different and quite subtle,
since there exist initial data that will lead to black holes in the evolution, see \cite{DR,AKR,A2b}.
If this happens, the spacetime will be geodesically incomplete.
However, even if black holes form, the solutions may still be global in (coordinate) time.
For instance, in so called Schwarzschild coordinates, it was shown in \cite{AKR} that there are initial data
that lead to the formation of black holes, where the solutions exist globally in time.
On the other hand, for initial data close to any non-trivial steady state, it is presently not known
if the corresponding solutions are global, which makes the stability problem quite intricate.
Let us also mention that for a trivial steady state, i.e., Minkowski space, global existence and stability
has been shown in \cite{FJS,LT}. Also if the discussion is restricted to linear stability and instability,
then rigorous results are available, see \cite{HR,HLR}.

In view of this discussion, one can expect that by obtaining a proof for the fact that static solutions 
minimize a particle number-Casimir functional over a certain set of functions, 
a new approach will be opened up for studying the non-linear stability problem.
Still though, and in contrast to the Newtonian case,
the path to a complete understanding of non-linear stability is shrouded in mystery.

The outline of the paper is as follows. In the next section 
we introduce the Einstein-Vlasov system. In Section \ref{patfpe} we derive the Euler-Lagrange equation 
discussed above and we formulate our main results. 
In Section \ref{ELsolv} we formulate the Euler-Lagrange equation as a fixed point problem 
for the energy density $\rho$ and we show existence of solutions to this equation. 
Having obtained a solution $\rho_\ast$ to the fixed point equation, 
we define in Section \ref{pot1} a density function $f_\ast$ that induces $\rho_\ast$ 
and that constitutes a static solution of the Einstein-Vlasov system. 
In Section \ref{rem_sect} the relation to stability is discussed. 
Finally, in Section \ref{app_sect} (which is an appendix) we collect properties 
of the Legendre transform and of some particular functions that are crucial for the argument. 


\section{The Einstein-Vlasov system}
 
\setcounter{equation}{0}

Below we use units such that $G=1$ and $c=1$, where $G$ is the gravitational constant 
and $c$ is the speed of light. For a function $g=g(t,r)$ we sometimes use the notation 
$g':=\partial_r g$ and $\dot{g}:=\partial_t g$. 
\smallskip

The metric of a spherically symmetric spacetime takes 
the following form in Schwarzschild coordinates:  
\[ ds^{2}=-e^{2\mu(t,r)}dt^{2}+e^{2\lambda(t,r)}dr^{2}
   +r^{2}(d\theta^{2}+\sin^{2}{\theta}\,d\varphi^{2}), \] 
where $r\geq 0,\,\theta\in [0,\pi],\,\varphi\in [0,2\pi]$ and $t\in \R$.
To ensure asymptotic flatness and a regular center, the following boundary conditions
are imposed:  
\[ \lim_{r\rightarrow\infty}\lambda(r)=\lim_{r\rightarrow\infty}\mu(r)
   =0=\lambda(t,0). \] 
We will now formulate the spherically symmetric Einstein-Vlasov system. We refer to \cite{A2,Rn1,Ren} 
for more information about this system and its derivation. The fundamental quantity that describes matter 
within the model is the density function $f$, which is defined on phase space. 
In this work we use two different coordinate systems on phase space which are standard in the literature, 
cf.~e.g.~\cite{Rn1}: either we write $f=f(t,r,w,l^2)$ or $f=f(x,v)=f(x_1,x_2,x_3,v_1,v_2,v_3)$, 
where these coordinates are related by 
\begin{eqnarray*}
   & & x=r(\sin\theta\cos\varphi, \sin\theta\sin\varphi, \cos\theta),\,\,\\
   & & w=\frac{x\cdot v}{r},\,\,\ell^2=|x\wedge v|^2. 
\end{eqnarray*}
The variables $w$ and $l^2$ can be thought of as the momentum in the radial
direction and the square of the angular momentum, respectively. 
Sometimes we will also use the notation
\[ \beta:=\ell^2. \] 
For later reference we note that $dx=4\pi r^2\,dr$, and that 
\begin{equation}\label{xvtransf}
   dx\,dv=8\pi^2\,dr\,d\ell\,\ell\,dw
   =4\pi^2\,dr\,d\ell^2\,dw.
\end{equation}
The Einstein-Vlasov system is given by the Einstein equations 
\begin{eqnarray} 
   & & e^{-2\lambda}(2r\lambda'-1)+1 = 8\pi r^2\,\rho, 
   \label{field1} 
   \\[1ex] & & e^{-2\lambda}(2r\mu'+1)-1 = 8\pi r^2\,p,
   \label{field2} 
   \\[1ex] & & \dot{\lambda} = -4\pi r\,e^{\lambda+\mu}\,j, 
   \nonumber 
   \\[1ex] & & e^{-2\lambda}\Big(\mu''+(\mu'-\lambda')\Big(\mu'+\frac{1}{r}\Big)\Big)
   -e^{-2\mu}\Big(\ddot{\lambda}+\dot{\lambda}(\dot{\lambda}-\dot{\mu})\Big)=8\pi\,p_T,
   \nonumber  
\end{eqnarray}   
together with the Vlasov equation 
\begin{eqnarray}
   & & \partial_t f+e^{\mu-\lambda}\,\frac{w}{\sqrt{1+w^2+\ell^2/r^2}}\,\partial_r f
   \nonumber
   \\ & & \hspace{1.7em} 
   -\,\Big(\dot{\lambda}\,w+e^{\mu-\lambda}\mu'\sqrt{1+w^2+\ell^2/r^2}
   -e^{\mu-\lambda}\,\frac{\ell^2}{r^3\sqrt{1+w^2+\ell^2/r^2}}\Big)\,\partial_w f=0. 
   \label{vlas2} 
\end{eqnarray}   
Here the non-vanishing components of the energy-momentum tensor are defined by
\begin{eqnarray}
   & & \rho(t, r)=\frac{\pi}{r^2}\int_{\R}\int_0^\infty\sqrt{1+w^2+\ell^2/r^2}
   \,f(t, r, w, \ell^2)\,d\ell^2\,dw,  
   \label{rho2} 
   \\[1ex] & & p(t, r)=\frac{\pi}{r^2}\int_{\R}\int_0^\infty\frac{w^2}{\sqrt{1+w^2+\ell^2/r^2}}
   \,f(t, r, w, \ell^2)\,d\ell^2\,dw,
   \label{p2} 
   \\[1ex] & & p_T(t, r)=\frac{\pi}{2r^4}\int_{\R}\int_0^\infty\frac{\ell^2}{\sqrt{1+w^2+\ell^2/r^2}}
   \,f(t, r, w, \ell^2)\,d\ell^2\,dw,
   \label{q2} 
   \\[1ex] & & j(t, r)=\frac{\pi}{r^2}\int_{\R}\int_0^\infty w\,f(t, r, w, \ell^2)\,d\ell^2\,dw. 
   \label{j2} 
\end{eqnarray}
The quantities $\rho, p, p_T$ and $j$ are the energy density, the radial pressure, 
the tangential pressure and the current, respectively. 
The equations above are not independent and typically one considers the reduced system (\ref{field1})-(\ref{field2}) 
together with (\ref{vlas2}) and (\ref{rho2})-(\ref{p2}). It is straightforward to show that a solution 
to the reduced system yields a solution to the full system, cf.~\cite{Rn1}. We will mainly consider 
static solutions in this work, and in the present coordinates such solutions are simply time independent. 

In the proofs we will use a couple of well-known consequences of the Einstein equations. 
First we note that equation (\ref{field1}) can be integrated to give 
\begin{equation}\label{mlam} 
   e^{-2\lambda(t, r)}=1-\frac{2m(t, r)}{r}, 
\end{equation}
where the mass function $m$ is defined by 
\begin{equation}\label{mform}  
   m(t, r)=4\pi\int_0^r s^2\rho(t, s)\,ds. 
\end{equation}
Moreover, equation (\ref{field2}) can be written in terms of $m$ and $\lambda$ as 
\begin{eqnarray}
   \mu(t, r)=-\int_r^\infty e^{2\lambda(t, s)}
   \Big(\frac{m(t, s)}{s^2}+4\pi s\,p(t, s)\Big)\,ds. 
   \label{mu_expl} 
\end{eqnarray}
If a density function $f$ is given we will denote by $\rho_f$, $p_f$, $m_f$, $\lambda_f$ and $\mu_f$ 
the functions defined by (\ref{rho2}), (\ref{p2}), ({\ref{mform}), (\ref{mlam}) and (\ref{mu_expl}), respectively.


\section{Preliminaries and main results}
\label{patfpe} 

\setcounter{equation}{0}

Some of the quantities below already appeared in \cite{Wol}. 
Let $\phi(s)=s_+^\sigma$ for $\sigma\in ]0, \frac{3}{2}[$ and $s_+=\max\{s, 0\}$. 
Define $\Phi(s)=\frac{1}{\sigma+1}\,s_+^{\sigma+1}$. Then the Legendre transform 
of $\Phi$ is calculated to be
\begin{equation}\label{ttl} 
   \hat{\Phi}(u)=\left\{\begin{array}{c@{\quad:\quad}l}
   \frac{\sigma}{\sigma+1}\,u^{1+1/\sigma} & u\in [0, \infty[
   \\[1ex] \infty & u\in ]-\infty, 0[\end{array}\right. ;
\end{equation} 
a few facts concerning Legendre transforms are recalled 
in Section \ref{Legrem}.
Next we define  
\begin{equation}\label{somal} 
   \Psi(\eps, r)=\int_0^\infty d\beta\int_{\R} dw\,\Phi(\alpha-\eps\sqrt{1+w^2+\beta/r^2})
\end{equation} 
for a fixed $\alpha>0$. We also let 
\begin{equation}\label{Gdef} 
   G(\eps)=\int_0^\infty\xi^2
   \,{(\alpha-\eps\sqrt{1+\xi^2})}_+^{\sigma+1}\,d\xi,\quad\eps\in\R. 
\end{equation} 
Using the change of variables 
$(w, y)=\xi(\cos\theta, \sin\theta)$, $|\det d(w, y)/d(\theta, \xi)|=\xi$, 
it is found that 
\begin{eqnarray}\label{PsiG}
   \Psi(\eps, r) & = & 
   \frac{1}{\sigma+1}\int_0^\infty d\beta\int_{\R} dw
   \,{(\alpha-\eps\sqrt{1+w^2+\beta/r^2})}_+^{\sigma+1}
   \nonumber
   \\ & = & \frac{2r^2}{\sigma+1}\int_0^\infty dy\,y\int_{\R} dw
   \,{(\alpha-\eps\sqrt{1+w^2+y^2})}_+^{\sigma+1}
   \nonumber
   \\ & = & \frac{2r^2}{\sigma+1}\int_0^\pi d\theta\sin\theta\int_0^\infty d\xi\,\xi^2
   \,{(\alpha-\eps\sqrt{1+\xi^2})}_+^{\sigma+1}
   \nonumber
   \\ & = & \frac{4r^2}{\sigma+1}\,G(\eps)
\end{eqnarray} 
for $G$ from (\ref{Gdef}). 
Another way to represent $\Psi$ is as follows. We have 
\begin{eqnarray}\label{Psi2} 
   \Psi(\eps, r) & = & \int_0^\infty d\beta\int_{\R} dw
   \,\frac{d}{dw}\,w\,\Phi(\alpha-\eps\sqrt{1+w^2+\beta/r^2})
   \nonumber
   \\ & = & \int_0^\infty d\beta\int_{\R} dw\,
   \Big[w\,\Phi(\alpha-\eps\sqrt{1+w^2+\beta/r^2})\Big]_{w=-\infty}^{w=\infty}
   \nonumber
   \\ & & +\,\eps\int_0^\infty d\beta\int_{\R} dw\,\frac{w^2}{\sqrt{1+w^2+\beta/r^2}}
   \,\Phi'(\alpha-\eps\sqrt{1+w^2+\beta/r^2})
   \nonumber
   \\ & = & \eps\int_0^\infty d\beta\int_{\R} dw\,\frac{w^2}{\sqrt{1+w^2+\beta/r^2}}
   \,\phi(\alpha-\eps\sqrt{1+w^2+\beta/r^2}). 
\end{eqnarray} 
Using (\ref{PsiG}), we calculate 
\[ \hat{\Psi}(u, r)=\sup_{\eps\in\R}\,\Big(\eps u-\frac{4r^2}{\sigma+1}\,G(\eps)\Big)
   =\frac{4r^2}{\sigma+1}\sup_{\eps\in\R}\,\Big(\frac{\sigma+1}{4r^2}\,\eps u-G(\eps)\Big)
   =\frac{4r^2}{\sigma+1}\,\hat{G}\Big(\frac{\sigma+1}{4r^2}\,u\Big), \] 
which can be re-expressed as  
\begin{equation}\label{rati2} 
   \hat{G}(u)=\frac{\sigma+1}{4r^2}\,\hat{\Psi}\Big(\frac{4r^2}{\sigma+1}\,u, r\Big). 
\end{equation} 
\bigskip 

Define the Lagrangian  
\[ L(r, q, p)=-\frac{r^{5/2}}{\sqrt{r-2q}}\,\inf_{\eps\in\R}
   \,\Big(G(\eps)+\eps\,\kappa\frac{p}{r^2}\Big)
   =\frac{r^{5/2}}{\sqrt{r-2q}}\,\hat{G}\Big(-\kappa\frac{p}{r^2}\Big) \]  
for $\kappa=\frac{\sigma+1}{16\pi^2}$. The derivatives are 
\begin{equation}\label{deL} 
   \frac{\partial L}{\partial q}=\frac{r^{5/2}}{(r-2q)^{3/2}}
   \,\hat{G}\Big(-\kappa\frac{p}{r^2}\Big),
   \quad\frac{\partial L}{\partial p}
   =-\frac{\kappa}{\sqrt{1-\frac{2q}{r}}}\,\hat{G}'\Big(-\kappa\frac{p}{r^2}\Big).
\end{equation}   
\medskip

Suppose now that there is a nice class of functions ${\cal A}\ni m$ 
such that  
\begin{equation}\label{calL_def} 
   {\cal L}(m)=\int_0^\infty L(r, m(r), m'(r))\,dr
\end{equation} 
is minimized over ${\cal A}$ by some nice function $m_\ast\in {\cal A}$ 
such that $\lim_{r\to 0}\frac{m_\ast(r)}{r}=0$, $m_\ast(\infty)=M$ and $\frac{2m_\ast(r)}{r}<1$. 
Define $\rho_\ast(r)=\frac{1}{4\pi r^2}\,m'_\ast(r)$. 
Then $4\pi\int_0^\infty r^2\,\rho_\ast(r)\,dr=\int_0^\infty m'_\ast(r)\,dr=m_\ast(\infty)-m_\ast(0)=M$, 
and one would expect that the Euler-Lagrange equation 
\[ \frac{d}{dr}\Big[\frac{\partial L}{\partial m'}\Big]=\frac{\partial L}{\partial m} \] 
is satisfied for $m_\ast$. Due to (\ref{deL}) this reads as 
\begin{equation}\label{ELmast} 
   -\,\kappa\,\frac{d}{dr}\bigg[\frac{1}{\sqrt{1-\frac{2m_\ast(r)}{r}}}
   \,\hat{G}'\Big(-\kappa\frac{m'_\ast(r)}{r^2}\Big)\bigg]
   =\frac{r^{5/2}}{(r-2m_\ast(r))^{3/2}}
   \,\hat{G}\Big(-\kappa\frac{m'_\ast(r)}{r^2}\Big).
\end{equation}
In the sequel we will study (\ref{ELmast}) in some more detail.  
\medskip

\begin{remark}\label{kleinl_intro}{\rm Let ${\cal G}(s)=\hat{G}(-s)$ and, 
for an appropriate function $m(r)=4\pi\int_0^r s^2\rho(s)\,ds$ such that $\frac{2m(r)}{r}<1$, 
put $\zeta(r)={\cal G}'(\kappa\frac{m'(r)}{r^2})={\cal G}'(4\pi\kappa\rho(r))$. 
Also define 
\[ l(r)=\kappa\,\frac{d}{dr}\bigg[\frac{1}{\sqrt{1-\frac{2m(r)}{r}}}
   \,\hat{G}'\Big(-\kappa\frac{m'(r)}{r^2}\Big)\bigg]
   +\frac{r^{5/2}}{(r-2m(r))^{3/2}}
   \,\hat{G}\Big(-\kappa\frac{m'(r)}{r^2}\Big), \] 
so that $m$ satisfies the Euler-Lagrange equations (\ref{ELmast}) 
on some interval $I\subset ]0, \infty[$ iff $l(r)=0$ for $r\in I$. 
Then $l$ can be rewritten as
\begin{equation}\label{kleinl_form} 
   l(r)=\frac{1}{\sqrt{1-\frac{2m(r)}{r}}}\,\bigg(
   -\kappa\,\zeta'(r)
   +\kappa\,\frac{m(r)}{r^2}\,\frac{1}{1-\frac{2m(r)}{r}}\,\zeta(r)
   -\frac{r}{1-\frac{2m(r)}{r}}\,G(-\zeta(r))\bigg),
\end{equation} 
cf.~\cite[eq.~(36), p.~221]{Wol} (which contains a misprint). 
To show this, using (\ref{dozu}), we have
\begin{eqnarray*} 
   l(r) & = & -\kappa\,\frac{d}{dr}\bigg[\frac{1}{\sqrt{1-\frac{2m(r)}{r}}}
   \,{\cal G}'\Big(\kappa\frac{m'(r)}{r^2}\Big)\bigg]
   +\frac{r^{5/2}}{(r-2m(r))^{3/2}}
   \,{\cal G}\Big(\kappa\frac{m'(r)}{r^2}\Big)
   \\ & = & -\kappa\,\frac{1}{\sqrt{1-\frac{2m(r)}{r}}}
   \,\frac{d}{dr}\,\Big[{\cal G}'\Big(\kappa\frac{m'(r)}{r^2}\Big)\Big]
   +\kappa\,\frac{1}{{(1-\frac{2m(r)}{r})}^{3/2}}\,\Big(\frac{m(r)}{r^2}-4\pi r\rho(r)\Big)
   \,{\cal G}'\Big(\kappa\frac{m'(r)}{r^2}\Big)
   \\ & & +\,\frac{r^{5/2}}{(r-2m(r))^{3/2}}
   \,{\cal G}\Big(\kappa\frac{m'(r)}{r^2}\Big)
   \\ & = & \frac{1}{\sqrt{1-\frac{2m(r)}{r}}}\,\Bigg(
   -\kappa\,\zeta'(r)
   +\kappa\,\frac{1}{1-\frac{2m(r)}{r}}\,\Big(\frac{m(r)}{r^2}-4\pi r\rho(r)\Big)
   \,{\cal G}'(4\pi\kappa\rho(r))
   \\ & & \hspace{7em} +\,\frac{r}{1-\frac{2m(r)}{r}}
   \,{\cal G}(4\pi\kappa\rho(r))\Bigg)
   \\ & = & \frac{1}{\sqrt{1-\frac{2m(r)}{r}}}\,\Bigg(
   -\kappa\,\zeta'(r)
   +\kappa\,\frac{1}{1-\frac{2m(r)}{r}}\,\frac{m(r)}{r^2}
   \,\zeta(r)
   \\ & & \hspace{7em} +\frac{r}{1-\frac{2m(r)}{r}}\Big[
   {\cal G}(4\pi\kappa\rho(r))-4\pi\kappa\rho(r)
   \,{\cal G}'(4\pi\kappa\rho(r))\Big]\Bigg)
   \\ & = & \frac{1}{\sqrt{1-\frac{2m(r)}{r}}}\,\Bigg(
   -\kappa\,\zeta'(r)
   +\kappa\,\frac{1}{1-\frac{2m(r)}{r}}\,\frac{m(r)}{r^2}\,\zeta(r)
   -\frac{r}{1-\frac{2m(r)}{r}}\,G(-{\cal G}'(4\pi\kappa\rho(r)))\Bigg), 
\end{eqnarray*} 
which yields (\ref{kleinl_form}). 

As a corollary, we note the following fact: if $l(r)=0$ 
for $r$ in some interval $I\subset ]0, \infty[$, 
then $\rho$ is strictly decreasing on $I$; see \cite[p.~221]{Wol}. 
Indeed, from (\ref{kleinl_form}) we get 
\[ \zeta'(r)=\frac{m(r)}{r^2}\,\frac{1}{1-\frac{2m(r)}{r}}\,\zeta(r)
   -\kappa^{-1}\frac{r}{1-\frac{2m(r)}{r}}\,G(-\zeta(r)),\quad r\in I. \] 
We have ${\cal G}'(s)<0$ for $s\in [0, \infty[$ by Lemma \ref{calGprops} 
and $G(\eps)>0$ for $\eps\in ]0, \alpha[$ by Lemma \ref{Geasy}. 
In addition, $-\zeta(r)=-{\cal G}'(4\pi\kappa\rho(r))\in ]0, \alpha[$, 
due to Lemma \ref{Gesasy}. Thus we obtain $\zeta'(r)<0$ for $r\in I$. 
If $r, \tilde{r}\in I$ are such that $r<\tilde{r}$, 
then ${\cal G}'(4\pi\kappa\rho(r))=\zeta(r)>\zeta(\tilde{r})={\cal G}'(4\pi\kappa\rho(\tilde{r}))$. 
Since ${\cal G}'$ is strictly increasing on $]0, \infty[$ 
due to Lemma \ref{calGprops}, it follows that $\rho(r)>\rho(\tilde{r})$.  
{\hfill$\diamondsuit$}
}
\end{remark}

Next we need to fix some constants. 
According to Lemmas \ref{Geasy}, \ref{Gesasy} and \ref{Gessasy} in the appendix 
(Section \ref{gs_sect}), there are $c_2>c_1>0$ 
(independent of $\alpha$, see Remark \ref{roi}) such that 
\begin{eqnarray} 
   & & c_1\,\alpha^{-3/2}\,(\alpha-\eps)^{5/2+\sigma}\le G(\eps)
   \le c_2\,\alpha^{-3/2}\,(\alpha-\eps)^{5/2+\sigma},  
   \label{baso1}
   \\[1ex] & & c_1\,\alpha^{-3/2}\,(\alpha-\eps)^{3/2+\sigma}\le |G'(\eps)|
   \le c_2\,\alpha^{-3/2}\,(\alpha-\eps)^{3/2+\sigma},  
   \label{baso9}
   \\[1ex] & & c_1\,\alpha^{-3/2}\,(\alpha-\eps)^{1/2+\sigma}\le G''(\eps)
   \le c_2\,\alpha^{-3/2}\,(\alpha-\eps)^{1/2+\sigma},  
   \label{baso2}
\end{eqnarray} 
all for $\eps\in [\frac{1}{\sqrt{2}}\,\alpha, \alpha]$. Due to 
Lemma \ref{calGprops} and (\ref{phca2}) there are further constants $c_3^\ast>0$ 
and $c_4>c_3>0$ (independent of $\alpha$) with the property that 
\begin{equation}\label{baso3} 
   c_3\,\alpha^{\frac{3}{3+2\sigma}}
   \,s^{-\frac{1+2\sigma}{3+2\sigma}}\le {\cal G}''(s)
   \le c_4\,\alpha^{\frac{3}{3+2\sigma}}
   \,s^{-\frac{1+2\sigma}{3+2\sigma}},
   \quad s\in ]0, s_0], 
\end{equation} 
for $s_0=2c_3^\ast\alpha^\sigma$. 

\begin{lemma}\label{niceprops} 
Suppose that $0\le\rho(r)\le\eta$, where $\eta\in ]0, 1]$ satisfies 
\begin{equation}\label{etacond} 
   \eta\le\min\Big\{\frac{c_3^\ast}{2\pi\kappa}, 
   \frac{1}{4\pi\kappa}\,\Big(\frac{\sqrt{2}}{\sqrt{2}-1}\Big)^{-\frac{3+2\sigma}{2}}
   \Big(\frac{3}{2}+\sigma\Big)^{-\frac{3+2\sigma}{2}} c_4^{-\frac{3+2\sigma}{2}}\Big\}\,\alpha^\sigma.
\end{equation} 
Then 
\begin{equation}\label{fristra} 
   c_5\,\alpha^{\frac{3}{3+2\sigma}}\rho(r)^{\frac{2}{3+2\sigma}}
   \le\alpha+\zeta(r)
   \le c_6\,\alpha^{\frac{3}{3+2\sigma}}\rho(r)^{\frac{2}{3+2\sigma}}
\end{equation} 
for $\zeta(r)={\cal G}'(4\pi\kappa\rho(r))$, defining
\[ c_5=c_3\,(4\pi\kappa)^{\frac{2}{3+2\sigma}}\,\frac{3+2\sigma}{2}
   \quad\mbox{and}\quad c_6=c_4\,(4\pi\kappa)^{\frac{2}{3+2\sigma}}\,\frac{3+2\sigma}{2}. \] 
In particular, we have 
\begin{equation}\label{minxiin} 
   -\zeta(r)\in\Big[\frac{1}{\sqrt{2}}\,\alpha, \alpha\Big]. 
\end{equation}
Furthermore, it holds that
\begin{eqnarray} 
   & & c_7\,\alpha^{\frac{3}{3+2\sigma}}\rho(r)^{\frac{5+2\sigma}{3+2\sigma}}
   \le G(-\zeta(r))
   \le c_8\,\alpha^{\frac{3}{3+2\sigma}}
   \rho(r)^{\frac{5+2\sigma}{3+2\sigma}},  
   \label{baso1eing}
   \\[1ex] & & c_7\,\alpha^{-\frac{3}{3+2\sigma}}
   \rho(r)^{\frac{1+2\sigma}{3+2\sigma}}\le G''(-\zeta(r))
   \le c_8\,\alpha^{-\frac{3}{3+2\sigma}}
   \rho(r)^{\frac{1+2\sigma}{3+2\sigma}},
   \label{baso2eing}
\end{eqnarray} 
for 
\[ c_7=\min\{c_5^{\frac{5+2\sigma}{2}}, c_5^{\frac{1+2\sigma}{2}}\}\,c_1
   \quad\mbox{and}\quad c_8=\max\{c_6^{\frac{5+2\sigma}{2}}, c_6^{\frac{1+2\sigma}{2}}\}\,c_2. \] 
\end{lemma} 
{\bf Proof\,:} Since $0\le 4\pi\kappa\rho(r)\le 4\pi\kappa\eta\le s_0$, 
for $s_0$ according to (\ref{baso3}), it follows from (\ref{baso3}) that 
\begin{eqnarray*} 
   \alpha+\zeta(r) & = & {\cal G}'(4\pi\kappa\rho(r))-{\cal G}'(0)
   =\int_0^{4\pi\kappa\rho(r)} {\cal G}''(\lambda)\,d\lambda
   \\ & \ge & c_3\,\alpha^{\frac{3}{3+2\sigma}}\int_0^{4\pi\kappa\rho(r)} 
   \lambda^{-\frac{1+2\sigma}{3+2\sigma}}\,d\lambda
   =c_5\,\alpha^{\frac{3}{3+2\sigma}}\rho(r)^{\frac{2}{3+2\sigma}}. 
\end{eqnarray*} 
The upper bound in (\ref{fristra}) can be shown in the same way, 
due to the upper bound on ${\cal G}''(s)$ from (\ref{baso3}). 
The lower bound for $-\zeta(r)$ in (\ref{minxiin}) is a consequence of (\ref{etacond}) and (\ref{fristra}), 
whereas the upper bound follow from (\ref{fristra}) and the fact that $\rho\geq 0$. 
Next, (\ref{baso1eing}) and (\ref{baso2eing}) are due to 
(\ref{baso1}) and (\ref{baso2}), combined with (\ref{fristra}), noticing that (\ref{minxiin}) holds. 
{\hfill$\Box$}\bigskip  

The main results in this work can now be stated. 

\begin{theorem}\label{EL_exi} For every fixed $\eta>0$ satisfying (\ref{etacond}) 
and for every $\alpha>0$ there exists a finite $R_\ast>0$ and a solution $m_\ast\in C^2([0, R_\ast])$ 
of the Euler-Lagrange equation (\ref{ELmast}) such that $0\le\frac{2m_\ast(r)}{r}<1$ 
for $r\in [0, R_\ast]$, $0<\rho_\ast(r)\le\eta$ for $r\in [0, R_\ast[$, 
$\rho_\ast(0)=\eta$, $\rho_\ast(R_\ast)=0$, and $\rho_\ast$ is strictly decreasing. 
\end{theorem} 

The solution $\rho_\ast$ induces a density function $f_\ast$ that is a static solution 
of the Einstein-Vlasov system. 

\begin{theorem}\label{mainthm1} 
Let $\phi(s)=s_+^\sigma$ for $\sigma\in ]0, \frac{3}{2}[$ be as above. 
For every $\alpha>0$ and $\eta>0$ sufficiently small (depending only on $\alpha$) 
there is a static solution to the Einstein-Vlasov system of the form 
\[ f_\ast(r, w, \beta)=\phi(\alpha-c e^{\mu(r)}\sqrt{1+w^2+\beta/r^2}) \]  
for $r\in [0, R]$, which is the support of the steady state; 
here $c>0$ is a suitable parameter that will be determined in the proof. 
Furthermore, the associated $\rho_\ast$ is a solution of the Euler-Lagrange equation (\ref{ELmast}). 
\end{theorem} 

\begin{remark}
{\rm Recall that the Euler-Lagrange equation (\ref{ELmast}) is associated 
with the functional (\ref{calL_def}), which in turn is closely related to the functional (\ref{pnC}) 
as shown in Section \ref{rem_sect}. Thus our result gives support to the claim that the static solutions 
constructed in Theorem \ref{mainthm1} are minimizers of the functional (\ref{pnC}) 
under reasonable constraints. However, a rigorous proof is yet missing. 
{\hfill$\diamondsuit$}
}
\end{remark}
\medskip


\section{Solving the Euler-Lagrange equation}
\label{ELsolv} 
\setcounter{equation}{0} 

Let $\eta$ be such that (\ref{etacond}) is verified. 
We consider the fixed point equation 
\begin{equation}\label{rho_eq} 
   \rho(r)=\eta-\frac{1}{4\pi\kappa^2}\int_0^r\,G''(-\zeta(s))\,\Big[\kappa\,\frac{m(s)}{s^2}\,
   \frac{1}{1-\frac{2m(s)}{s}}\,(-\zeta(s))+\frac{s}{1-\frac{2m(s)}{s}}\,G(-\zeta(s))\Big]\,ds. 
\end{equation} 
Note that this is not an integrated ODE in the standard form, 
since the dependence of $m$ on $\rho$ is $m(r)=4\pi\int_0^r s^2\rho(s)\,ds$. Equation (\ref{rho_eq}) is 
obtained from the equation $l(r)=0$, where $l(r)$ is given by equation (\ref{kleinl_form}), by using the properties of the Legendre transform, as specified in Section \ref{Legrem}. 
\smallskip 

We start with some auxiliary observations regarding the right-hand side of (\ref{rho_eq}). 

\begin{lemma}\label{sitr}
(a) Let $R>0$ and $\rho\in C([0, R])$ be such that $\rho(r)>0$ and $\frac{2m(r)}{r}<1$ 
for $r\in [0, R]$. Define 
\begin{equation}\label{Irho_def} 
   I_\rho(r)=G''(-\zeta(r))\,\Big[\kappa\,\frac{m(r)}{r^2}\,
   \frac{1}{1-\frac{2m(r)}{r}}\,(-\zeta(r))+\frac{r}{1-\frac{2m(r)}{r}}\,G(-\zeta(r))\Big],
   \quad r\in [0, R],
\end{equation}  
to be the integrand in (\ref{rho_eq}), where $\frac{m(r)}{r^2}$ and $\frac{m(r)}{r}$ 
are taken to be zero at $r=0$. Then $I_\rho(r)>0$ for $r\in ]0, R]$. 
\smallskip 

\noindent 
(b) If $\rho\in C([0, R])$ is a solution to (\ref{rho_eq}) on some interval $[0, R]$, 
then in fact $\rho$ is continuously differentiable on $[0, R]$ 
and satisfies $\rho'(r)=-\frac{1}{4\pi\kappa^2}\,I_\rho(r)$, 
so that $\rho'(r)<0$ for $r\in ]0, R]$. 
\smallskip 

\noindent 
(c) There is a constant $c_9>0$ with the following property. 
Let $R>0$ and $\rho\in C([0, R])$ be such that $0<\rho(r)\le\eta$ 
and $\frac{2m(r)}{r}\le\frac{1}{2}$ for $r\in [0, R]$. Then 
\[ I_\rho(r)\le c_9 (\alpha^{\frac{2\sigma}{3+2\sigma}}
   \eta^{\frac{4(1+\sigma)}{3+2\sigma}}+\eta^2)\,R. \] 
\smallskip 

\noindent 
(d) There is a constant $c_{10}>0$ with the following property.
Let $R>0$ and $\rho, \tilde{\rho}\in C([0, R])$ be such that $0<\rho(r)\le\eta$, 
$0<\tilde{\rho}(r)\le\eta$ and $\frac{2m(r)}{r}<1$, $\frac{2\tilde{m}(r)}{r}<1$  
for $r\in [0, R]$. Then one can find a constant $\hat{C}=\hat{C}(\rho, \tilde{\rho})>0$ such that 
\begin{eqnarray}\label{juba}  
   \lefteqn{|I_{\tilde{\rho}}(r)-I_\rho(r)|} 
   \nonumber
   \\ & \le & c_{10}\,\hat{C}\Big(\alpha^{\frac{\sigma(1+2\sigma)}{3+2\sigma}}
   \,(\eta+\alpha^{-\frac{2\sigma}{3+2\sigma}}\eta^{\frac{5+2\sigma}{3+2\sigma}})
   +\eta^{\frac{4(1+\sigma)}{3+2\sigma}}
   +\alpha^{\sigma}\,\eta^{\frac{1+2\sigma}{3+2\sigma}}\Big)
   \,R\,|\tilde{\rho}(r)-\rho(r)|
   \nonumber
   \\ & & +\,c_{10}\,\hat{C}\Big(R^2(\eta^2+\alpha^{\frac{2\sigma}{3+2\sigma}}
   \,\eta^{\frac{4(1+\sigma)}{3+2\sigma}})
   +\alpha^{\frac{2\sigma}{3+2\sigma}}\,\eta^{\frac{1+2\sigma}{3+2\sigma}}\Big)
   \,\int_0^r |\tilde{\rho}(s)-\rho(s)|\,ds. 
\end{eqnarray}
\smallskip 

\noindent 
(e) There is a constant $\bar{C}=\bar{C}(\alpha, \eta)>0$ with the following property.
Let $R>0$ and $\rho, \tilde{\rho}\in C([0, R])$ be such that $\frac{\eta}{2}\le\rho(r)\le\eta$, 
$\frac{\eta}{2}\le\tilde{\rho}(r)\le\eta$ and $\frac{2m(r)}{r}\le\frac{1}{2}$, $\frac{2\tilde{m}(r)}{r}\le\frac{1}{2}$  
for $r\in [0, R]$. Then 
\begin{equation}\label{juba12}  
   |I_{\tilde{\rho}}(r)-I_\rho(r)|
   \le\bar{C} R\,|\tilde{\rho}(r)-\rho(r)|
   +\bar{C}(R^2+1)\,\int_0^r |\tilde{\rho}(s)-\rho(s)|\,ds. 
\end{equation}
\end{lemma} 
{\bf Proof\,:} (a) Since $G''(\eps)>0$ for $\eps\in ]0, \alpha[$ by Lemma \ref{Gessasy} 
and $-\zeta(r)=-{\cal G}'(4\pi\kappa\rho(r))\in ]0, \alpha[$ according to Lemma \ref{calGprops}, 
we deduce that $G''(-\zeta(r))>0$. Also $G(\eps)>0$ for $\eps\in ]0,\alpha[$ 
by Lemma \ref{Geasy}, so that $G(-\zeta(r))>0$. 
As a consequence, we get $I_\rho(r)>0$ for $r\in ]0, R]$. 
(b) This follows from (a). 
(c) From $0\le\frac{2m(r)}{r}\le\frac{1}{2}$ we obtain 
$\frac{1}{1-\frac{2m(r)}{r}}\in [1, 2]$ for $r\in [0, R]$. 
Moreover, $0\le m(r)=4\pi\int_0^r s^2\rho(s)\,ds\le\frac{4\pi}{3}\,\eta\,r^3$, 
so that $\frac{m(r)}{r^2}\le\frac{4\pi}{3}\,\eta\,r\le\frac{4\pi}{3}\,\eta\,R$ 
for $r\in [0, R]$. Thus we deduce from (\ref{baso2eing}) and (\ref{minxiin}) that 
\begin{eqnarray*} 
   0 & \le & \kappa\,G''(-\zeta(r))\,\frac{m(r)}{r^2}\,
   \frac{1}{1-\frac{2m(r)}{r}}\,(-\zeta(r))
   \\ & \le & \kappa\,c_8\,\alpha^{-\frac{3}{3+2\sigma}}
   \rho(r)^{\frac{1+2\sigma}{3+2\sigma}}\,\frac{8\pi}{3}\,\eta\,R\,\alpha 
   \\ & \le & \frac{8\pi}{3}\,\kappa\,c_8\,\alpha^{\frac{2\sigma}{3+2\sigma}}
   \eta^{\frac{4(1+\sigma)}{3+2\sigma}}\,R.   
\end{eqnarray*} 
Similarly, by (\ref{baso2eing}) and (\ref{baso1eing}), 
\begin{eqnarray*} 
   0 & \le & G''(-\zeta(s))\,\frac{s}{1-\frac{2m(s)}{s}}\,G(-\zeta(s))
   \\ & \le & c_8\,\alpha^{-\frac{3}{3+2\sigma}}
   \rho(r)^{\frac{1+2\sigma}{3+2\sigma}}\,2R\,c_8\,\alpha^{\frac{3}{3+2\sigma}}
   \rho(r)^{\frac{5+2\sigma}{3+2\sigma}}
   \\ & \le & 2c_8^2\,\eta^2\,R, 
\end{eqnarray*}
so that altogether 
\[ I_\rho(r)\le\Big(\frac{8\pi}{3}\,\kappa\,c_8\,\alpha^{\frac{2\eta}{3+2\sigma}}
   \eta^{\frac{4(1+\sigma)}{3+2\sigma}}
   +2c_8^2\,\eta^2\Big)\,R, \] 
which shows that we can take $c_9=\max\{\frac{8\pi}{3}\,\kappa\,c_8, 2c_8^2\}$. 
(d) Let $\delta>0$ be such that $\delta\le\rho(r)\le\eta$, 
$\delta\le\tilde{\rho}(r)\le\eta$ and $\frac{2m(r)}{r}\le 1-\delta$, 
$\frac{2\tilde{m}(r)}{r}\le 1-\delta$ for $r\in [0, R]$. 
Then there is $q\in [\frac{1}{\sqrt{2}}, 1[$ (that can be calculated from $\delta$) 
so that $-\zeta(r)=-{\cal G}'(4\pi\kappa\rho(r))\in [\frac{1}{\sqrt{2}}\alpha, q\alpha]$ 
as well as $-\tilde{\zeta}(r)=-{\cal G}'(4\pi\kappa\tilde{\rho}(r))
\in [\frac{1}{\sqrt{2}}\alpha, q\alpha]$; for this, also recall (\ref{minxiin}). 
Given this $q$, let $C_q>0$ denote the associated constant from Corollary \ref{GessLip}. 
We have 
\begin{eqnarray*} 
   \lefteqn{|I_{\tilde{\rho}}(r)-I_\rho(r)|} 
   \\ & \le & |G''(-\tilde{\zeta}(r))-G''(-\zeta(r))|\,\Big[\kappa\,\frac{\tilde{m}(r)}{r^2}\,
   \frac{1}{1-\frac{2\tilde{m}(r)}{r}}\,(-\tilde{\zeta}(r))+\frac{r}{1-\frac{2\tilde{m}(r)}{r}}
   \,G(-\tilde{\zeta}(r))\Big]
   \\ & & +\,\kappa\,|G''(-\zeta(r))|
   \,\Big|\frac{\tilde{m}(r)}{r^2}\,
   \frac{1}{1-\frac{2\tilde{m}(r)}{r}}\,(-\tilde{\zeta}(r))
   -\frac{m(r)}{r^2}\,\frac{1}{1-\frac{2m(r)}{r}}\,(-\zeta(r))\Big|
   \\ & & +\,|G''(-\zeta(r))|
   \,\Big|\frac{r}{1-\frac{2\tilde{m}(r)}{r}}\,G(-\tilde{\zeta}(r))
   -\frac{r}{1-\frac{2m(r)}{r}}\,G(-\zeta(r))\Big|
   \\ & \le & |G''(-\tilde{\zeta}(r))-G''(-\zeta(r))|\,\Big[\kappa\,\frac{\tilde{m}(r)}{r^2}\,
   \frac{1}{1-\frac{2\tilde{m}(r)}{r}}\,(-\tilde{\zeta}(r))+\frac{r}{1-\frac{2\tilde{m}(r)}{r}}
   \,G(-\tilde{\zeta}(r))\Big]
   \\ & & +\,\kappa\,|G''(-\zeta(r))|\,\frac{\tilde{m}(r)}{r^2}
   \,\frac{1}{1-\frac{2\tilde{m}(r)}{r}}\,|\tilde{\zeta}(r)-\zeta(r)|
   \\ & & +\,2\kappa\,|G''(-\zeta(r))|\,\frac{\tilde{m}(r)}{r^3}\,(-\zeta(r))
   \,\frac{|\tilde{m}(r)-m(r)|}{(1-\frac{2\tilde{m}(r)}{r})(1-\frac{2m(r)}{r})}
   \\ & & +\,\kappa\,|G''(-\zeta(r))|\,(-\zeta(r))\,\frac{1}{1-\frac{2m(r)}{r}}
   \,\frac{|\tilde{m}(r)-m(r)|}{r^2}
   \\ & & +\,|G''(-\zeta(r))|
   \,\frac{r}{1-\frac{2\tilde{m}(r)}{r}}\,|G(-\tilde{\zeta}(r))-G(-\zeta(r))|
   \\ & & +\,2\,|G''(-\zeta(r))|\,|G(-\zeta(r))|
   \frac{|\tilde{m}(r)-m(r)|}{(1-\frac{2\tilde{m}(r)}{r})(1-\frac{2m(r)}{r})}
   \\ & =: & T_1+T_2+T_3+T_4+T_5+T_6. 
\end{eqnarray*} 
Henceforth $C>0$ will denote a generic constant (that is independent 
of $R$, $\alpha$, $\eta$, $\delta$). To bound $T_1$, by (\ref{baso3}) and (\ref{etacond}) we have 
\begin{eqnarray}\label{diffzeta}  
   |\tilde{\zeta}(r)-\zeta(r)| 
   & = & |{\cal G}'(4\pi\kappa\tilde{\rho}(r))-{\cal G}'(4\pi\kappa\rho(r))|
   \nonumber
   \\ & \le & 4\pi\kappa\,(\max_{s\in [4\pi\kappa\delta, 4\pi\kappa\eta]}\,{\cal G}''(s))
   \,|\tilde{\rho}(r)-\rho(r)| 
   \nonumber
   \\ & \le & 4\pi\kappa\,c_4\,\alpha^{\frac{3}{3+2\sigma}}
   \,(4\pi\kappa\delta)^{-\frac{1+2\sigma}{3+2\sigma}}
   \,|\tilde{\rho}(r)-\rho(r)|
   \nonumber
   \\ & \le & C\,\alpha^{\frac{3}{3+2\sigma}}
   \,\delta^{-\frac{1+2\sigma}{3+2\sigma}}
   \,|\tilde{\rho}(r)-\rho(r)|. 
\end{eqnarray} 
Therefore, using Corollary \ref{GessLip}, 
\begin{eqnarray*} 
   |G''(-\tilde{\zeta}(r))-G''(-\zeta(r))|
   & \le & C_q\,\alpha^{-(2-\sigma)}\,|\tilde{\zeta}(r)-\zeta(r)|
   \\ & \le & C C_q\,\alpha^{-\frac{(\sigma+1)(3-2\sigma)}{3+2\sigma}}
   \,\delta^{-\frac{1+2\sigma}{3+2\sigma}}
   \,|\tilde{\rho}(r)-\rho(r)|. 
\end{eqnarray*} 
Moreover,  
\begin{equation}\label{tilm} 
   \tilde{m}(r)=4\pi\int_0^r s^2\rho(s)\,ds\le\frac{4\pi}{3}\,\eta\,r^3
\end{equation} 
and 
\begin{equation}\label{hxi} 
   G(-\tilde{\zeta}(r))\le c_8\,\alpha^{\frac{3}{3+2\sigma}}
   \tilde{\rho}(r)^{\frac{5+2\sigma}{3+2\sigma}}
   \le c_8\,\alpha^{\frac{3}{3+2\sigma}}
   \eta^{\frac{5+2\sigma}{3+2\sigma}}
\end{equation} 
by (\ref{baso1eing}). Taking these estimates together, if follows that 
\[ T_1\le C C_q\,\alpha^{\frac{\sigma(1+2\sigma)}{3+2\sigma}}
   \,\delta^{-\frac{4(1+\sigma)}{3+2\sigma}}
   \,(\eta+\alpha^{-\frac{2\sigma}{3+2\sigma}}
   \eta^{\frac{5+2\sigma}{3+2\sigma}})\,R\,|\tilde{\rho}(r)-\rho(r)|. \] 
For $T_2$, we use (\ref{baso2eing}) to obtain 
\begin{equation}\label{kais} 
   |G''(-\zeta(r))|\le c_8\,\alpha^{-\frac{3}{3+2\sigma}}
   \tilde{\rho}(r)^{\frac{1+2\sigma}{3+2\sigma}}
   \le c_8\,\alpha^{-\frac{3}{3+2\sigma}}
   \eta^{\frac{1+2\sigma}{3+2\sigma}}.
\end{equation} 
Therefore (\ref{tilm}) and (\ref{diffzeta}) yield 
\[ T_2\le C\,\delta^{-\frac{4(1+\sigma)}{3+2\sigma}}
   \,\eta^{\frac{4(1+\sigma)}{3+2\sigma}}
   \,R\,|\tilde{\rho}(r)-\rho(r)|. \] 
Next, 
\begin{equation}\label{sowu} 
   |\tilde{m}(r)-m(r)|\le 4\pi\int_0^r s^2\,|\tilde{\rho}(s)-\rho(s)|\,ds
   \le 4\pi r^2\int_0^r |\tilde{\rho}(s)-\rho(s)|\,ds,
\end{equation} 
and thus from (\ref{kais}) and (\ref{tilm}) we get 
\[ T_3\le C\,\alpha^{\frac{2\sigma}{3+2\sigma}}
   \,\delta^{-2}\,\eta^{\frac{4(1+\sigma)}{3+2\sigma}}
   \,R^2\int_0^r |\tilde{\rho}(s)-\rho(s)|\,ds. \] 
For $T_4$ we can argue in a similar way, and due to $r^{-2}|\tilde{m(r)}-m(r)|
\le 4\pi\int_0^r |\tilde{\rho}(s)-\rho(s)|\,ds$ we also have 
\[ T_4\le C\,\alpha^{\frac{2\sigma}{3+2\sigma}}
   \,\delta^{-1}\,\eta^{\frac{1+2\sigma}{3+2\sigma}}
   \,\int_0^r |\tilde{\rho}(s)-\rho(s)|\,ds. \] 
Regarding $T_5$, by using (\ref{baso9}) and (\ref{diffzeta}) we deduce 
\begin{eqnarray*} 
   |G(-\tilde{\zeta}(r))-G(-\zeta(r))|
   & \le & (\max_{\eps\in [\frac{1}{\sqrt{2}}\alpha, q\alpha]} |G'(\eps)|)
   \,|\tilde{\zeta}(r)-\zeta(r)|
   \\ & \le & C\alpha^{-3/2}\,(\max_{\eps\in [\frac{1}{\sqrt{2}}\alpha, q\alpha]}(\alpha-\eps)^{3/2+\sigma})
   \,\alpha^{\frac{3}{3+2\sigma}}
   \,\delta^{-\frac{1+2\sigma}{3+2\sigma}}
   \,|\tilde{\rho}(r)-\rho(r)|
   \\ & \le & C\alpha^{\frac{2\sigma^2+3\sigma+3}{3+2\sigma}}
   \,\delta^{-\frac{1+2\sigma}{3+2\sigma}}
   \,|\tilde{\rho}(r)-\rho(r)|. 
\end{eqnarray*} 
Hence, as a consequence of (\ref{kais}), 
\[ T_5\le C\alpha^{\sigma}\,\delta^{-\frac{4(1+\sigma)}{3+2\sigma}}\,\eta^{\frac{1+2\sigma}{3+2\sigma}}
   \,R\,|\tilde{\rho}(r)-\rho(r)|. \] 
Lastly, the bound 
\[ T_6\le C\,\delta^{-2}\eta^2\,R^2\int_0^r |\tilde{\rho}(s)-\rho(s)|\,ds \]
is again due to (\ref{kais}), (\ref{hxi}) and (\ref{sowu}). 
Taking together all the above estimates on $T_1, \ldots, T_6$, we arrive at 
\begin{eqnarray}\label{kido}  
   \lefteqn{|I_{\tilde{\rho}}(r)-I_\rho(r)|} 
   \nonumber
   \\ & \le & C\delta^{-\frac{4(1+\sigma)}{3+2\sigma}}\Big(C_q\,\alpha^{\frac{\sigma(1+2\sigma)}{3+2\sigma}}
   \,(\eta+\alpha^{-\frac{2\sigma}{3+2\sigma}}\eta^{\frac{5+2\sigma}{3+2\sigma}})
   +\eta^{\frac{4(1+\sigma)}{3+2\sigma}}
   +\alpha^{\sigma}\,\eta^{\frac{1+2\sigma}{3+2\sigma}}\Big)
   \,R\,|\tilde{\rho}(r)-\rho(r)|   
   \nonumber
   \\ & & +\,C\Big(\delta^{-2} R^2(\eta^2+\alpha^{\frac{2\sigma}{3+2\sigma}}\,\eta^{\frac{4(1+\sigma)}{3+2\sigma}})
   +\alpha^{\frac{2\sigma}{3+2\sigma}}\,\delta^{-1}\,\eta^{\frac{1+2\sigma}{3+2\sigma}}\Big)
   \,\int_0^r |\tilde{\rho}(s)-\rho(s)|\,ds,  
\end{eqnarray} 
after some regrouping of terms, which yields (\ref{juba}). 
(e) This follows from an inspection of the argument for (d) and from (\ref{kido}). 
We can take $\delta=\min\{\frac{\eta}{2}, \frac{1}{2}\}$ everywhere, 
and since $-\zeta(r)=-{\cal G}'(4\pi\kappa\rho(r))\in [\frac{1}{\sqrt{2}}\alpha, -{\cal G}'(2\pi\kappa\eta)]$, 
we need to take $q=-\alpha^{-1}\,{\cal G}'(2\pi\kappa\eta)$ 
when we apply Corollary \ref{GessLip} to determine $C_q$. Note that with some efforts 
the constant $\bar{C}=\bar{C}(\alpha, \eta)$ could be calculated explicitly, since $C_q$ is explicit. 
{\hfill$\Box$}\bigskip 

First we have to deal with uniqueness of solutions to (\ref{rho_eq}).  

\begin{lemma}\label{uniqsol} 
Let $R>0$ and let $\rho, \tilde{\rho}\in C([0, R])$ 
be solutions of (\ref{rho_eq}) such that $0<\rho(r)\le\eta$, 
$0<\tilde{\rho}(r)\le\eta$ and $\frac{2m(r)}{r}<1$, $\frac{2\tilde{m}(r)}{r}<1$  
for $r\in [0, R]$. Then $\rho(r)=\tilde{\rho}(r)$ for $r\in [0, R]$. 
\end{lemma} 
{\bf Proof\,:} Due to (\ref{juba}) in Lemma \ref{sitr}(d) there is a constant 
$\tilde{C}=\tilde{C}(\rho, \tilde{\rho}, R, \alpha, \eta)>0$ such that 
\[ |I_{\tilde{\rho}}(r)-I_\rho(r)|
   \le\tilde{C}\Big(|\tilde{\rho}(r)-\rho(r)|+\int_0^r |\tilde{\rho}(s)-\rho(s)|\,ds\Big),
   \quad r\in [0, R], \] 
owing to (\ref{juba}). By (\ref{rho_eq}) this yields 
\begin{eqnarray*} 
   |\tilde{\rho}(r)-\rho(r)| 
   & \le & \frac{1}{4\pi\kappa^2}\int_0^r |I_{\tilde{\rho}}(s)-I_\rho(s)|\,ds
   \\ & \le & \frac{\tilde{C}}{4\pi\kappa^2}\int_0^r
   \Big(|\tilde{\rho}(s)-\rho(s)|+\int_0^s |\tilde{\rho}(\tau)-\rho(\tau)|\,d\tau\Big)\,ds
\end{eqnarray*} 
for $r\in [0, R]$. Denote $\Delta(r)=\max_{s\in [0, r]} |\tilde{\rho}(s)-\rho(s)|$. 
Since $\Delta$ is monotone, it follows that 
\[ |\tilde{\rho}(r)-\rho(r)| 
   \le\frac{\tilde{C}}{4\pi\kappa^2}\int_0^r (\Delta(s)+s\Delta(s))\,ds
   \le\frac{\tilde{C}}{4\pi\kappa^2}\,(1+R)\int_0^r\Delta(s)\,ds \] 
for $r\in [0, R]$. Fix $r\in [0, R]$ and $\bar{r}\in [0, r]$. Then 
\[ |\tilde{\rho}(\bar{r})-\rho(\bar{r})|
   \le\frac{\tilde{C}}{4\pi\kappa^2}\,(1+R)\int_0^{\bar{r}}\Delta(s)\,ds
   \le\frac{\tilde{C}}{4\pi\kappa^2}\,(1+R)\int_0^r\Delta(s)\,ds \]
yields 
\[ \Delta(r)\le\frac{\tilde{C}}{4\pi\kappa^2}\,(1+R)\int_0^r\Delta(s)\,ds, \]
and hence Gronwall's inequality applies. 
{\hfill$\Box$}\bigskip   

Next comes the local existence of a solution to (\ref{rho_eq}). 

\begin{lemma}\label{locexsol} 
Let $c_9>0$ be the constant from Lemma \ref{sitr}(c) 
and denote by $\bar{C}=\bar{C}(\alpha, \eta)>0$ the constant from Lemma \ref{sitr}(e). 
Suppose that $R\in ]0, 1]$ is so small that 
\begin{equation}\label{Rconds}
   (\alpha^{\frac{2\sigma}{3+2\sigma}}
   \eta^{\frac{1+2\sigma}{3+2\sigma}}+\eta)\,R^2\le\frac{2\pi\kappa^2}{c_9}, 
   \quad\eta\,R^2\le\frac{3}{16\pi}
   \quad\mbox{and}\quad\bar{C} R^2\le\frac{2\pi\kappa^2}{3}. 
\end{equation}
Then (\ref{rho_eq}) has a (unique) solution $\rho\in C([0, R])$ so that 
$\rho(0)=\eta$, $\frac{\eta}{2}\le\rho(r)\le\eta$ and $\frac{2m(r)}{r}\le\frac{1}{2}$ 
for $r\in [0, R]$.  
\end{lemma} 
{\bf Proof\,:} This is a standard application of Banach's fixed point theorem 
on the closed set 
\[ D=\Big\{\rho\in C([0, R]): \rho(0)=\eta, 
   \frac{\eta}{2}\le\rho(r)\le\eta\,\,\mbox{and}\,\,\frac{2m(r)}{r}\le\frac{1}{2}, 
   r\in [0, R]\Big\} \] 
in the Banach space $X=C([0, R])$, and for the operator 
\[ F: D\to D,\quad (F\rho)(r)=\eta-\frac{1}{4\pi\kappa^2}\int_0^r I_\rho(s)\,ds, \] 
with $I_\rho$ given by (\ref{Irho_def}). First we will show that $F:D\to D$ is well-defined. 
For, let $\rho\in D$. Clearly $(F\rho)(0)=0$, and since $I_\rho(s)>0$ for $s\in ]0, R]$ 
by Lemma \ref{sitr}(a), we also have $(F\rho)(r)\le\eta$ for $r\in [0, R]$.  
What concerns the lower bound, from Lemma \ref{sitr}(c) we recall that 
\[ I_\rho(r)\le c_9 (\alpha^{\frac{2\sigma}{3+2\sigma}}
   \eta^{\frac{4(1+\sigma)}{3+2\sigma}}+\eta^2)\,R \] 
for $r\in [0, R]$. Hence 
\[ (F\rho)(r)\ge\eta-\frac{1}{4\pi\kappa^2}\,c_9 (\alpha^{\frac{2\sigma}{3+2\sigma}}
   \eta^{\frac{4(1+\sigma)}{3+2\sigma}}+\eta^2)\,R^2\ge\frac{\eta}{2} \] 
for $r\in [0, R]$, where we have used (\ref{Rconds}). 
Next, for the mass generated by $F\rho$ we obtain 
\[ \frac{2}{r}\,4\pi\int_0^r s^2 (F\rho)(s)\,ds
   \le\frac{8\pi}{3r}\,\eta\,r^3\le\frac{8\pi}{3}\,\eta\,R^2\le\frac{1}{2} \] 
for $r\in [0, R]$, once again by (\ref{Rconds}). Therefore we have seen that 
indeed $F(D)\subset D$ is verified. For the contraction property, 
if $\rho, \tilde{\rho}\in D$ and $r\in [0, R]$, then 
\[ |(F\rho)(r)-(F\tilde{\rho})(r)|
   \le\frac{1}{4\pi\kappa^2}\int_0^r |I_{\tilde{\rho}}(s)-I_\rho(s)|\,ds. \]  
Due to (\ref{juba12}) from Lemma \ref{sitr}(e) we have  
\begin{eqnarray*} 
   |I_{\tilde{\rho}}(s)-I_\rho(s)|
   & \le & \bar{C} R\,|\tilde{\rho}(s)-\rho(s)|
   +\bar{C}(R^2+1)\,\int_0^s |\tilde{\rho}(\tau)-\rho(\tau)|\,d\tau
   \\ & \le & \bar{C} R\,(R^2+2)\,{\|\tilde{\rho}-\rho\|}_X, 
\end{eqnarray*} 
so that 
\[ {\|F\rho-F\tilde{\rho}\|}_X
   \le\frac{1}{4\pi\kappa^2}\,\bar{C} R^2\,(R^2+2)\,{\|\tilde{\rho}-\rho\|}_X
   \le\frac{1}{2}\,{\|\tilde{\rho}-\rho\|}_X, \]  
once again using (\ref{Rconds}). Thus $F$ is a contraction. 
{\hfill$\Box$}\bigskip   


We are now in a position to prove Theorem \ref{EL_exi}. 
\smallskip

\noindent
{\bf Proof of Theorem \ref{EL_exi}\,:} Consider the maximal solution $\rho_\ast$ of (\ref{rho_eq}) 
such that $0<\rho_\ast(r)\le\eta$ and $\frac{2m_\ast(r)}{r}<1$ holds, 
which is defined on some interval $[0, R_\ast[$, where $0<R_\ast\le\infty$. 
Such a maximal solution does exist (by Zorn's lemma), since there is some 
solution with these properties on some interval $[0, R]$ by Lemma \ref{locexsol}, 
and solutions with these properties are unique by Lemma \ref{uniqsol}. 
According to Lemma \ref{sitr}(b) we also know that $\rho_\ast\in C^1([0, R_\ast[)$ 
satisfies $\rho'_\ast(r)<0$ for $r\in [0, R_\ast[$. 
Let $\zeta_\ast(r)={\cal G}'(4\pi\kappa\rho_\ast(r))$ be as before.
From the choice of $\eta$ and (\ref{minxiin}) in Lemma \ref{niceprops} 
we infer that $-\zeta_\ast(r)\in [\frac{1}{\sqrt{2}}\alpha, \alpha]$ for $r\in [0, R_\ast[$. 
Differentiating (\ref{rho_eq}), we obtain 
\begin{equation}\label{bablu} 
   \rho'_\ast(r)=-\frac{1}{4\pi\kappa^2}\,G''(-\zeta_\ast(r))\,\Big[\kappa\,\frac{m_\ast(r)}{r^2}\,
   \frac{1}{1-\frac{2m_\ast(r)}{r}}\,(-\zeta_\ast(r))
   +\frac{r}{1-\frac{2m_\ast(r)}{r}}\,G(-\zeta_\ast(r))\Big],
   \quad r\in [0, R_\ast[. 
\end{equation} 
As a consequence, 
\begin{eqnarray}\label{alpsp}  
   \kappa\,\zeta'_\ast(r) & = & 4\pi\kappa^2\,{\cal G}''(4\pi\kappa\rho_\ast(r))\,\rho'_\ast(r)
   \nonumber
   \\ & = & {\cal G}''(4\pi\kappa\rho_\ast(r))\,
   \,G''(-\zeta_\ast(r))\,\Big[\kappa\,\frac{m_\ast(r)}{r^2}\,
   \frac{1}{1-\frac{2m_\ast(r)}{r}}\,\zeta_\ast(r)
   -\frac{r}{1-\frac{2m_\ast(r)}{r}}\,G(-\zeta_\ast(r))\Big]. 
   \nonumber \\ & & 
\end{eqnarray} 
For $s=4\pi\kappa\rho_\ast(r)$ and $u=-s=-4\pi\kappa\rho_\ast(r)$ 
we have $\eps(u)=\hat{G}'(u)$ by (\ref{derLeg}). Hence ${\cal G}(s)=\hat{G}(-s)$ 
shows that $\eps(-s)=\eps(u)=\hat{G}'(u)=-{\cal G}'(-u)=-{\cal G}'(s)=-\zeta_\ast(r)$. 
Therefore we can apply (\ref{nwt}), and (\ref{alpsp}) simplifies to  
\begin{equation}\label{zetaode} 
   \zeta'_\ast(r)=\frac{m_\ast(r)}{r^2}\,
   \frac{1}{1-\frac{2m_\ast(r)}{r}}\,\zeta_\ast(r)
   -\kappa^{-1}\frac{r}{1-\frac{2m_\ast(r)}{r}}\,G(-\zeta_\ast(r)),
   \quad r\in [0, R_\ast[.
\end{equation} 

First we are going to show that $R_\ast<\infty$ is verified, 
and for this we will use an argument similar to the one in \cite[Prop.~4.3]{Wol}. 
Suppose that we have $R_\ast=\infty$. Then (\ref{rho_eq}) holds for $r\in [0, \infty[$, 
and moreover $0<\rho_\ast(r)\le\eta$ and $\frac{2m_\ast(r)}{r}<1$ for $r\in [0, \infty[$. 
We apply the change of variables $s=r^3$, $r=s^{1/3}$, and define 
\[ \hat{\rho}(s)=\rho_\ast(r)=\rho_\ast(s^{1/3}),
   \quad \hat{m}(s)=m_\ast(r)=m_\ast(s^{1/3}), 
   \quad \hat{\zeta}(s)=\zeta_\ast(r)=\zeta_\ast(s^{1/3}). \] 
Note that always $2\hat{m}(s)<s^{1/3}$ due to $\frac{2m_\ast(r)}{r}<1$. In addition, 
\[ \hat{m}'(s)=\frac{1}{3}\,s^{-2/3}\,m'_\ast(s^{1/3})
   =\frac{1}{3}\,s^{-2/3}\,4\pi s^{2/3}\,\rho_\ast(s^{1/3})
   =\frac{4\pi}{3}\,\rho_\ast(s^{1/3}), \] 
and thus 
\[ \hat{m}''(s)=\frac{4\pi}{9}\,s^{-2/3}\,\rho_\ast'(s^{1/3})<0 \]     
for $s\in ]0, \infty[$, which implies that $\hat{m}$ is concave 
on $[0, \infty[$. Hence $(s-s_0)\hat{m}'(s)\le\hat{m}(s)-\hat{m}(s_0)$ 
for $0\le s_0\le s$, so that $\hat{m}(0)=0$ yields 
\begin{equation}\label{sty} 
   \hat{\rho}(s)=\frac{3}{4\pi}\,\hat{m}'(s)\le\frac{3}{4\pi}\,\frac{\hat{m}(s)}{s},
   \quad s\in ]0, \infty[.
\end{equation} 
Moreover, in the variable $s$, (\ref{zetaode}) is found to read as 
\begin{equation}\label{zetaodes} 
   \hat{\zeta}'(s)
   =\frac{\hat{m}(s)}{3s}\,\frac{1}{(s^{1/3}-2\hat{m}(s))}\,\hat{\zeta}(s)
   -\frac{1}{3\kappa}\,\frac{1}{(s^{1/3}-2\hat{m}(s))}\,G(-\hat{\zeta}(s)),
   \quad s\in ]0, \infty[.
\end{equation} 
Since $\hat{\zeta}(s)<0$ and $\hat{m}(s)\ge 0$ we have 
\[ \frac{\hat{m}(s)}{3s}\,\frac{1}{(s^{1/3}-2\hat{m}(s))}\,\hat{\zeta}(s)
   \le\frac{\hat{m}(s)}{3s^{4/3}}\,\hat{\zeta}(s) \] 
and hence $G(-\hat{\zeta}(s))\ge 0$ in conjunction with (\ref{zetaodes}) leads to 
\begin{equation}\label{cons1} 
   \hat{\zeta}'(s)\le\frac{\hat{m}(s)}{3s^{4/3}}\,\hat{\zeta}(s),
   \quad s\in ]0, \infty[.
\end{equation}
Similarly, from 
\[ -\frac{1}{3\kappa}\,\frac{1}{(s^{1/3}-2\hat{m}(s))}\,G(-\hat{\zeta}(s))
   \le -\frac{1}{3\kappa}\,\frac{1}{s^{1/3}}\,G(-\hat{\zeta}(s)) \] 
together with (\ref{zetaodes}) we find that 
\begin{equation}\label{cons2} 
   \hat{\zeta}'(s)\le -\frac{1}{3\kappa}\,\frac{1}{s^{1/3}}\,G(-\hat{\zeta}(s)),
   \quad s\in ]0, \infty[, 
\end{equation} 
holds. Those two differential inequalities (\ref{cons1}) and (\ref{cons2}) 
for $\hat{\zeta}$ can be used as follows. Due to (\ref{baso1}) 
and $-\hat{\zeta}(s)\in [\frac{1}{\sqrt{2}}\alpha, \alpha]$ we have 
\[ G(-\hat{\zeta}(s))\ge c_1\,\alpha^{-3/2}\,(\alpha+\hat{\zeta}(s))^{5/2+\sigma}. \] 
Therefore (\ref{cons2}) yields 
\begin{eqnarray}\label{luga}  
   \frac{d}{ds}\,(\alpha+\hat{\zeta}(s))^{-(3/2+\sigma)}
   & = & -\Big(\frac{3+2\sigma}{2}\Big)\,(\alpha+\hat{\zeta}(s))^{-(5/2+\sigma)}\,\hat{\zeta}'(s)
   \nonumber
   \\ & \ge & \Big(\frac{3+2\sigma}{6\kappa}\Big)\,\frac{1}{s^{1/3}}
   \,(\alpha+\hat{\zeta}(s))^{-(5/2+\sigma)}\,G(-\hat{\zeta}(s))
   \nonumber 
   \\ & \ge & c_1\Big(\frac{3+2\sigma}{6\kappa}\Big)\,\alpha^{-3/2}\,\frac{1}{s^{1/3}},
   \quad s\in ]0, \infty[. 
\end{eqnarray} 
Now we need to distinguish two cases. (i) If $\hat{m}(s)\le 1$ for all $s\in [0, \infty[$, 
then $\hat{M}=\lim_{s\to\infty}\hat{m}(s)\in ]0, 1]$ does exist, since $\hat{m}$ is increasing.  
Here we take $s_\ast>0$ so large that 
\begin{equation}\label{sast1} 
   \hat{m}(s)\ge\frac{\hat{M}}{2}\quad\mbox{and}\quad
   s^{\frac{3-2\sigma}{6}}>\frac{3\cdot 20^{\frac{3+2\sigma}{2}}}{4\pi}\,c_6^{\frac{3+2\sigma}{2}}
   \alpha^{-\sigma}\,\Big(\frac{2}{\hat{M}}\Big)^{\frac{3+2\sigma}{2}}
\end{equation} 
holds for all $s\ge s_\ast$; recall that $3-2\sigma>0$. 
(ii) If $\hat{m}(s)>1$ for $s\ge s_0$ (with an appropriate $s_0>0$) is verified, 
then we determine $s_\ast\ge s_0$ such that 
\begin{equation}\label{sast2} 
   s^{\frac{3-2\sigma}{6}}>\frac{3\cdot 20^{\frac{3+2\sigma}{2}}}{4\pi}
   \,c_6^{\frac{3+2\sigma}{2}}\alpha^{-\sigma}
\end{equation} 
for $s\ge s_\ast$. Consider $s\ge s_\ast$. Integrating (\ref{luga}), we arrive at 
\[ (\alpha+\hat{\zeta}(s))^{-(3/2+\sigma)}-(\alpha+\hat{\zeta}(s_\ast))^{-(3/2+\sigma)}
   \ge c_{11}\,\alpha^{-3/2}\,(s^{2/3}-s_\ast^{2/3}),
   \quad s\in [s_\ast, \infty[, \] 
where $c_{11}=c_1(\frac{3+2\sigma}{4\kappa})$. This can be recast as
\begin{equation}\label{cons3} 
   0\le\alpha+\hat{\zeta}(s)
   \le (\alpha+\hat{\zeta}(s_\ast))
   \Big[1+c_{11}\,\alpha^{-3/2}
   \,(\alpha+\hat{\zeta}(s_\ast))^{3/2+\sigma}\,(s^{2/3}-s_\ast^{2/3})\Big]^{-\frac{2}{3+2\sigma}},
   \quad s\in [s_\ast, \infty[,
\end{equation} 
Next, $-\hat{\zeta}(s)\ge\frac{1}{\sqrt{2}}\,\alpha\ge\frac{1}{2}\,\alpha$. 
Thus, owing to (\ref{cons1}),
\[ \hat{\zeta}'(s)\le -\frac{\alpha}{6}\,\frac{\hat{m}(s)}{s^{4/3}},
   \quad s\in ]0, \infty[. \] 
Hence for $s\ge s_\ast$ and $\Delta>0$ the monotonicity of $\hat{m}$ implies that 
\begin{equation}\label{cons4} 
   \hat{\zeta}(s+\Delta)-\hat{\zeta}(s)
   \le -\frac{\alpha}{6}\int_s^{\Delta+s}\frac{\hat{m}(\tau)}{\tau^{4/3}}\,d\tau
   \le -\frac{\alpha}{2}\,\hat{m}(s)\,(s^{-1/3}-(\Delta+s)^{-1/3}).
\end{equation} 
Applying (\ref{cons4}) to $s=2s_\ast$ and $\Delta=s_\ast$, we find 
\[ \alpha+\hat{\zeta}(3s_\ast)
   \le\alpha+\hat{\zeta}(2s_\ast)-\frac{\alpha}{2}\,\frac{\hat{m}(2s_\ast)}{s_\ast^{1/3}}
   \,\Big(\frac{1}{2^{1/3}}-\frac{1}{3^{1/3}}\Big)
   \le\alpha+\hat{\zeta}(2s_\ast)-\frac{\alpha}{20}\,\frac{\hat{m}(2s_\ast)}{s_\ast^{1/3}}. \] 
Thus taking $s=2s_\ast$ in (\ref{cons3}) to bound $\alpha+\hat{\zeta}(2s_\ast)$ 
on the right-hand side, we get
\begin{eqnarray}\label{gostr} 
   \alpha+\hat{\zeta}(3s_\ast)
   & \le & (\alpha+\hat{\zeta}(s_\ast))\Big[1+c_{11}\,\alpha^{-3/2}
   \,(\alpha+\hat{\zeta}(s_\ast))^{3/2+\sigma}\,s_\ast^{2/3}(2^{2/3}-1)\Big]^{-\frac{2}{3+2\sigma}}
   -\frac{\alpha}{20}\,\frac{\hat{m}(2s_\ast)}{s_\ast^{1/3}}
   \nonumber
   \\ & \le & (\alpha+\hat{\zeta}(s_\ast))\Big[1+\frac{c_{11}}{2}\,\alpha^{-3/2}
   \,(\alpha+\hat{\zeta}(s_\ast))^{3/2+\sigma}\,s_\ast^{2/3}\Big]^{-\frac{2}{3+2\sigma}}
   -\frac{\alpha}{20}\,\frac{\hat{m}(2s_\ast)}{s_\ast^{1/3}}. 
\end{eqnarray} 
From (\ref{fristra}) and $\rho_\ast(r)>0$ for $r\in [0, \infty[$ we deduce that actually
\[ \alpha+\hat{\zeta}(3s_\ast)=\alpha+\zeta_\ast((3s_\ast)^{1/3})
   \ge c_5\,\alpha^{\frac{3}{3+2\sigma}}\rho_\ast((3s_\ast)^{1/3})^{\frac{2}{3+2\sigma}}>0 \] 
for the left-hand side of (\ref{gostr}). However, we are going to show that 
\begin{equation}\label{frapr} 
   s_\ast^{\frac{3+2\sigma}{6}}(\alpha+\hat{\zeta}(s_\ast))^{\frac{3+2\sigma}{2}}
   <\Big(\frac{\alpha}{20}\Big)^{\frac{3+2\sigma}{2}}
   \,\Big[1+\frac{c_{11}}{2}\,\alpha^{-3/2}
   \,(\alpha+\hat{\zeta}(s_\ast))^{3/2+\sigma}\,s_\ast^{2/3}\Big]
   \,\hat{m}(2s_\ast)^{\frac{3+2\sigma}{2}},
\end{equation} 
and then the desired contradiction will be reached. To establish (\ref{frapr}), 
we use (\ref{fristra}) and (\ref{sty}) to estimate 
\begin{eqnarray}\label{juhol}  
   s_\ast^{\frac{3+2\sigma}{6}}(\alpha+\hat{\zeta}(s_\ast))^{\frac{3+2\sigma}{2}}
   & = & s_\ast^{\frac{3+2\sigma}{6}}(\alpha+\zeta_\ast(s_\ast^{1/3}))^{\frac{3+2\sigma}{2}}
   \le c_6^{\frac{3+2\sigma}{2}}\alpha^{3/2}\,s_\ast^{\frac{3+2\sigma}{6}}\rho_\ast(s_\ast^{1/3}) 
   =c_6^{\frac{3+2\sigma}{2}}\alpha^{3/2}\,s_\ast^{\frac{3+2\sigma}{6}}\hat{\rho}(s_\ast)
   \nonumber
   \\ & \le & \frac{3}{4\pi}\,c_6^{\frac{3+2\sigma}{2}}\alpha^{3/2}\,s_\ast^{\frac{3+2\sigma}{6}}
   \,\frac{\hat{m}(s_\ast)}{s_\ast}
   =\frac{3}{4\pi}\,c_6^{\frac{3+2\sigma}{2}}\alpha^{3/2}\,s_\ast^{-\frac{3-2\sigma}{6}}
   \,\hat{m}(s_\ast). 
\end{eqnarray} 
At this point we have to come back to the cases (i) and (ii) that we distinguished above, 
and which led to different choices of $s_\ast$. Case (i): $\hat{m}(s)\le 1$ for all $s\in [0, \infty[$. 
Then by (\ref{juhol}) and (\ref{sast1}), 
\begin{eqnarray*} 
   s_\ast^{\frac{3+2\sigma}{6}}(\alpha+\hat{\zeta}(s_\ast))^{\frac{3+2\sigma}{2}}
   & \le & \frac{3}{4\pi}\,c_6^{\frac{3+2\sigma}{2}}\alpha^{3/2}\,s_\ast^{-\frac{3-2\sigma}{6}}
   \\ & < & \Big(\frac{\alpha}{20}\Big)^{\frac{3+2\sigma}{2}}\,\Big(\frac{\hat{M}}{2}\Big)^{\frac{3+2\sigma}{2}}
   \le\Big(\frac{\alpha}{20}\Big)^{\frac{3+2\sigma}{2}}\,\hat{m}(2s_\ast)^{\frac{3+2\sigma}{2}}, 
\end{eqnarray*} 
which proves (\ref{frapr}). Case (ii): $\hat{m}(s)>1$ for $s\ge s_0$. 
Then by (\ref{juhol}) and the monotonicity of $\hat{m}$, 
\begin{eqnarray*} 
   s_\ast^{\frac{3+2\sigma}{6}}(\alpha+\hat{\zeta}(s_\ast))^{\frac{3+2\sigma}{2}}
   & \le & \frac{3}{4\pi}\,c_6^{\frac{3+2\sigma}{2}}\alpha^{3/2}\,s_\ast^{-\frac{3-2\sigma}{6}}
   \,\hat{m}(2s_\ast)<\Big(\frac{\alpha}{20}\Big)^{\frac{3+2\sigma}{2}}
   \,\hat{m}(2s_\ast)^{\frac{3+2\sigma}{2}}, 
\end{eqnarray*} 
where we used (\ref{sast2}), $\hat{m}(2s_\ast)>1$ as well as $\frac{3+2\sigma}{2}>1$. 
Therefore (\ref{frapr}) holds in all cases, which in turn proves that $R_\ast=\infty$ 
is impossible. 

Therefore we must have $R_\ast<\infty$. Since both $\rho_\ast$ and $m_\ast$ 
are monotone, the limits $\rho_\ast(R_\ast):=\lim_{r\to R_\ast}\rho_\ast(r)\in [0, \eta]$ 
and $m_\ast(R_\ast)=\lim_{r\to R_\ast} m_\ast(r)$ do exist; 
note that $\frac{2m_\ast(r)}{r}<1$ for $r\in [0, R_\ast[$ implies that $m_\ast$ 
is bounded on $[0, R_\ast]$ and $m_\ast(R_\ast)\le R_\ast/2$. 

If $\rho_\ast(R_\ast)>0$ and $\frac{2m_\ast(R_\ast)}{R_\ast}<1$, 
then an argument similar to the one in the proof of Lemma \ref{locexsol} 
shows that the solution $\rho_\ast$ can be slightly extended 
beyond $R_\ast$ to a solution $\tilde{\rho}$ on some interval $[0, R_\ast+\eps[$ 
such that $0<\tilde{\rho}(r)\le\eta$ and $\frac{2\tilde{m}(r)}{r}<1$ 
for $r\in [0, R_\ast+\eps[$, which however contradicts the fact that $\rho$ is maximal. 

The next case to consider is $R_\ast<\infty$ and $\frac{2m_\ast(R_\ast)}{R_\ast}=1$. 
Let $r_\ast\in ]0, R_\ast[$ be such that $\frac{2m_\ast(r_\ast)}{r_\ast}\ge\frac{1}{2}$. 
From (\ref{zetaode}) we obtain 
\[ \frac{d}{dr}\,\ln (-\zeta_\ast(r))=\frac{\zeta'_\ast(r)}{\zeta_\ast(r)}
   =\frac{m_\ast(r)}{r^2}\,
   \frac{1}{1-\frac{2m_\ast(r)}{r}}
   +\kappa^{-1}\frac{r}{1-\frac{2m_\ast(r)}{r}}\,\frac{G(-\zeta_\ast(r))}{(-\zeta_\ast(r))} \] 
for $r\in [0, R_\ast[$, whence $\frac{G(-\zeta_\ast(r))}{(-\zeta_\ast(r))}\ge 0$ 
in conjunction with (\ref{minxiin}) yields
\[ \alpha\ge (-\zeta_\ast(R_\ast))
   \ge (-\zeta_\ast(r_\ast))\,e^{\int_{r_\ast}^{R_\ast}\frac{m_\ast(r)}{r(r-2m_\ast(r))}\,dr}
   \ge \frac{1}{\sqrt{2}}\,\alpha\,e^{\int_{r_\ast}^{R_\ast}\frac{m_\ast(r)}{r(r-2m_\ast(r))}\,dr}. \]  
Therefore 
\[ \frac{1}{R_\ast}\int_{r_\ast}^{R_\ast}\frac{m_\ast(r)}{r-2m_\ast(r)}\,dr
   \le\int_{r_\ast}^{R_\ast}\frac{m_\ast(r)}{r(r-2m_\ast(r))}\,dr
   \le\ln\sqrt{2}. \] 
Since $m_\ast$ is monotone, we get $m_\ast(r)\ge m_\ast(r_\ast)\ge\frac{r_\ast}{4}$ 
for $r\in [r_\ast, R_\ast]$. It follows that 
\begin{equation}\label{intbd} 
   \int_{r_\ast}^{R_\ast}\frac{dr}{R_\ast-2m_\ast(r)}
   \le\int_{r_\ast}^{R_\ast}\frac{dr}{r-2m_\ast(r)}
   \le 4\ln\sqrt{2}\,\,\frac{R_\ast}{r_\ast}.
\end{equation} 
Due to $m'_\ast(r)=4\pi r^2\rho_\ast(r)>0$ for $r\in ]0, R_\ast[$, 
the function $[r_\ast, R_\ast]\ni r\mapsto s(r)=m_\ast(r)\in [m_\ast(r_\ast), m_\ast(R_\ast)]
=[s_\ast, S_\ast]$ is invertible. Thus changing variables as $s=m_\ast(r)$, $ds=m'_\ast(r)\,dr$, 
we deduce from (\ref{intbd})  
\[ \frac{1}{4\pi R_\ast^2}\int_{s_\ast}^{S_\ast}\frac{ds}{\rho_\ast(r(s))(R_\ast-2s)}
   \le\int_{s_\ast}^{S_\ast}\frac{ds}{4\pi r^2\rho_\ast(r(s))(R_\ast-2s)}
   \le 4\ln\sqrt{2}\,\,\frac{R_\ast}{r_\ast}. \] 
As $S_\ast=m_\ast(R_\ast)=R_\ast/2$ we get 
\[ \int_{s_\ast}^{S_\ast}\frac{ds}{\rho_\ast(r(s))(S_\ast-s)}
   \le 32\pi\ln\sqrt{2}\,\,\frac{R_\ast^3}{r_\ast}. \]
Now $0<\rho_\ast(r)\le\eta$ for $r\in [0, R_\ast]$. Therefore we arrive at the contradiction 
\[ \int_{s_\ast}^{S_\ast}\frac{ds}{S_\ast-s}
   \le 32\pi\ln\sqrt{2}\,\,\frac{R_\ast^3}{r_\ast}\,\eta. \]

If we summarize the above cases so far, we can deduce that in fact we must have 
$R_\ast<\infty$, $\rho_\ast(R_\ast)=0$ and $\frac{2m_\ast(R_\ast)}{R_\ast}<1$. 
Since we know that (\ref{zetaode}) holds, this implies that $l(r)=0$ for $l$ from (\ref{kleinl_form}). 
According to Remark \ref{kleinl_intro}, this means that $m_\ast$ verifies 
the Euler-Lagrange equation. For the last assertion about the monotonicity 
of $\rho_\ast$, this follows from Lemma \ref{sitr}(b). 
{\hfill$\Box$}\bigskip  

\begin{remark} 
{\rm (a) In \cite{Wol} it is asserted that for all (sufficiently small) ADM masses $M$ 
there is a static solution of this mass $M$. Note that so far this is not included 
in our results, since both the end of the support $R_\ast$ of $\rho_\ast$ 
and $M=4\pi\int_0^{R_\ast} r^2\rho_\ast(r)\,dr$ are not explicit. 
\smallskip 

\noindent 
(b) One might wonder what goes wrong with the whole argument in the massless case. 
Typically massless (small) solutions are not expected to have compact support, cf.~\cite{AFT}. 
Also if $G$ from (\ref{Gdef}) would be replaced by 
\begin{equation}\label{Gdef2} 
   \tilde{G}(\eps)=\int_0^\infty\xi^2
   \,{(\alpha-\eps\xi)}_+^{\sigma+1}\,d\xi,\quad\eps\in\R, 
\end{equation}
then the estimates for $G$ and $\tilde{G}$ differ drastically and the arguments in this work break down. 
\smallskip

\noindent
(c) The bound $2m(r)/r<1$ obtained in the proof can be improved once we have a static solution. 
Indeed, in \cite{A1} it is shown that $2m(r)/r<8/9$ when $p\geq 0$ and $p+2p_T\leq \rho$. 
Both conditions hold in this case. 
{\hfill$\diamondsuit$}
}
\end{remark}
\medskip


\section{Proof of Theorem \ref{mainthm1}}
\label{pot1} 
\setcounter{equation}{0} 

We will show that a solution to the Euler-Lagrange equation (\ref{ELmast}) 
yields a static solution. First we define a set of functions $\Xi_r$ that is essential 
for our approach. Let $r>0$ be fixed and put
\[ \Xi_r=\{\chi=\chi(w, \beta)\in L^{1+1/\sigma}(\R\times [0, \infty[): 
   \chi\ge 0, \chi(w, \beta)=0\,\,\mbox{for}\,\,\sqrt{1+w^2+\beta/r^2}\ge 2\}. \] 
Then $\Xi_r\subset L^{1+1/\sigma}(\R\times [0, \infty[)$ 
is closed and convex; the number ``2'' is chosen such that the support is sufficiently large. 
The size of the support is dictated by the estimate of $\zeta$ in Lemma \ref{niceprops}. 
Now consider the functionals 
\begin{equation}\label{HF-def} 
   H(\chi)=\int_0^\infty d\beta\int_{\R} dw\,(\hat{\Phi}(\chi)-\alpha\chi),
   \quad F(\chi, r, a)=\int_0^\infty d\beta\int_{\R} dw\,
   \sqrt{1+w^2+\beta/r^2}\,\chi
   -\frac{r^2}{\pi}\,a,
\end{equation} 
for $r>0$, functions $\chi=\chi(w, \beta)\in\Xi_r$ and $a>0$. 

\begin{lemma}\label{wibirk} For $r>0$ and $a>0$ 
there is a unique function $\chi=\chi(w, \beta; r, a)\in\Xi_r$ such that 
\begin{equation}\label{lslem} 
   \inf_{\chi\in\Xi_r}\,\{H(\chi): F(\chi, r, a)=0\}
   =H(\chi(\cdot, \cdot; r, a))
\end{equation}  
and 
\begin{equation}\label{F0} 
   F(\chi(\cdot, \cdot; r, a), r, a)=0.
\end{equation}
For $r>0$ and $a=0$ this unique function is $\chi(w, \beta; r, 0)=0$. 
\end{lemma} 
{\bf Proof\,:} Define $I=\inf_{\chi\in\Xi_r}\,\{H(\chi): F(\chi, r, a)=0\}$ 
and let $\chi\in\Xi_r$ be such that $F(\chi, r, a)=0$. Then 
\begin{eqnarray*} 
   H(\chi) & = & \int_0^\infty d\beta\int_{\R} dw\,
   \Big(\frac{\sigma}{\sigma+1}\,\chi^{1+1/\sigma}-\alpha\chi\Big)
   \ge -\alpha\int_0^\infty d\beta\int_{\R} dw\,\chi
   \\ & \ge & -\alpha\int_0^\infty d\beta\int_{\R} dw\,\sqrt{1+w^2+\beta/r^2}\,\chi
   = -\alpha\,\frac{r^2}{\pi}\,a.  
\end{eqnarray*} 
Thus $I\ge -\alpha\frac{r^2}{\pi} a$ is finite. 
The function $x\mapsto x^{1+1/\sigma}$ is strictly convex and the part 
of $H$ containing $\alpha\chi$ is linear. Therefore $H$ itself is strictly convex. 
In addition, $H$ is continuous on $\Xi_r\subset L^{1+1/\sigma}(\R\times [0, \infty[)$, 
since the $L^1(\R\times [0, \infty[)$-norm on $\Xi_r$ 
can be controlled in terms of the $L^{1+1/\sigma}(\R\times [0, \infty[)$-norm, 
due to the supports of functions in $\Xi_r$ being contained in 
$\{\sqrt{1+w^2+\beta/r^2}\le 2\}$. 
Let $(\chi_j)\subset\Xi_r$ be a minimizing sequence for $I$, 
i.e., we have $\lim_{j\to\infty} H(\chi_j)=I$ and $F(\chi_j, r, a)=0$ 
for $j\in\N$. Then 
\begin{eqnarray*} 
   \frac{\sigma}{\sigma+1}\int_0^\infty d\beta\int_{\R} dw\,\chi_j^{1+1/\sigma}
   & = & H(\chi_j)+\alpha\int_0^\infty d\beta\int_{\R} dw\,\chi_j
   \\ & \le & H(\chi_j)+\alpha\int_0^\infty d\beta\int_{\R} dw\,\sqrt{1+w^2+\beta/r^2}\,\chi_j
   \\ & = & H(\chi_j)+\alpha\,\frac{r^2}{\pi}\,a
\end{eqnarray*}  
shows that $(\chi_j)\subset L^{1+1/\sigma}(\R\times [0, \infty[)$ 
is bounded. Hence we may assume that $\chi_j\rightharpoonup\chi$ weakly 
in $L^{1+1/\sigma}(\R\times [0, \infty[)$ as $j\to\infty$. 
Since $\Xi_r$ is weakly closed, we infer that $\chi\in\Xi_r$, 
and moreover $H(\chi)\le\liminf_{j\to\infty} H(\chi_j)=I$. 
For $j\in\N$ we have  
\begin{equation}\label{upto} 
   \int_0^\infty d\beta\int_{\R} dw\,{\bf 1}_{\{\sqrt{1+w^2+\beta/r^2}\le 2\}}
   \sqrt{1+w^2+\beta/r^2}\,\chi_j=\frac{r^2}{\pi}\,a.
\end{equation}  
Since $\varphi\in L^{1+\sigma}(\R\times [0, \infty[)$ 
for $\varphi(w, \beta)={\bf 1}_{\{\sqrt{1+w^2+\beta/r^2}\le 2\}}
\sqrt{1+w^2+\beta/r^2}$, we may pass to the limit $j\to\infty$ in (\ref{upto}) 
to obtain 
\[ \int_0^\infty d\beta\int_{\R} dw\,{\bf 1}_{\{\sqrt{1+w^2+\beta/r^2}\le 2\}}
   \sqrt{1+w^2+\beta/r^2}\,\chi=\frac{r^2}{\pi}\,a, \] 
which means that $F(\chi, r, a)=0$, and hence $\chi$ is a minimizer. 

For the asserted uniqueness, note that this follows from 
the strict convexity of $H$ and the convexity of the set 
$\{\chi\in\Xi_r: F(\chi, r, a)=0\}$. In fact, if $\chi, \tilde{\chi}\in\Xi_r$ 
are minimizers, then $F(\chi, r, a)=F(\tilde{\chi}, r, a)=0$ 
implies that $F((\chi+\tilde{\chi})/2, r, a)=0$ 
and $(\chi+\tilde{\chi})/2\in\Xi_r$. If we had $\chi\neq\tilde{\chi}$, 
then $I\le H((\chi+\tilde{\chi})/2)<H(\chi)/2+H(\tilde{\chi})/2=I$, 
which is impossible. 
{\hfill$\Box$}\bigskip  

Now we are in a position to introduce the desired static solution. 
Let $R_\ast>0$, $m_\ast$ and $\rho_\ast$ be given by Proposition \ref{EL_exi} 
and put $M=m_{\ast}(R_\ast)$. Also recall that by construction 
$0\le\frac{2m_\ast(r)}{r}<1$ as well as $0\le\rho_\ast(r)\le\eta$ 
for $r\in [0, R_\ast]$. Another preliminary observation is that 
\begin{eqnarray*}
   -{\cal G}'(4\pi\kappa\eta)
   & = & \alpha-({\cal G}'(4\pi\kappa\eta)-{\cal G}'(0))
   =\alpha-\int_0^{4\pi\kappa\eta} {\cal G}''(\lambda)\,d\lambda
   \\ & \ge & \alpha-c_4\,\alpha^{\frac{3}{3+2\sigma}}\int_0^{4\pi\kappa\eta}
   \,\lambda^{-\frac{1+2\sigma}{3+2\sigma}}\,d\lambda
   =\alpha-\Big(\frac{3}{2}+\sigma\Big) c_4 (4\pi\kappa)^{\frac{2}{3+2\sigma}}
   \,\alpha^{\frac{3}{3+2\sigma}}\,\eta^{\frac{2}{3+2\sigma}}
   \ge\frac{1}{\sqrt{2}}\,\alpha,   
\end{eqnarray*} 
where we have used (\ref{baso3}) and the second condition on $\eta$ in (\ref{etacond}). Define 
\begin{equation}\label{iap} 
   f_\ast(r, w, \beta)=\chi(w, \beta; r, \rho_\ast(r)),
   \quad\lambda_\ast(r)=-\frac{1}{2}\ln\Big(1-\frac{2m_\ast(r)}{r}\Big),
\end{equation}
where $\chi$ is the function obtained in Lemma \ref{wibirk}.  
Then $m'_\ast(r)=4\pi r^2\rho_\ast(r)$ and $e^{-2\lambda_\ast(r)}=1-\frac{2m_\ast(r)}{r}$ 
shows that (\ref{field1}) is satisfied. Also $\lim_{r\to 0}\frac{m_\ast(r)}{r}=0$ 
and $m_\ast(\infty)=M$ imply that $\lambda_\ast(0)=\lambda_\ast(\infty)=0$. 
To specify $\mu_\ast$, note first that 
\begin{equation}\label{pastform} 
   p_\ast(r)=\int\frac{w^2}{\sqrt{1+v^2}}\,f_\ast\,dv
   =\frac{\pi}{r^2}\int_0^\infty d\beta\int_{\R} dw\,
   \frac{w^2}{\sqrt{1+w^2+\beta/r^2}}\,f_\ast(r, w, \beta)
\end{equation} 
can already be calculated from $f_\ast$. Thus it is possible to introduce $\mu_\ast$ 
by requiring that
\begin{equation}\label{muast_intro} 
   \mu'_\ast(r)=e^{2\lambda_\ast(r)}\Big(\frac{m_\ast(r)}{r^2}+4\pi rp_\ast(r)\Big),
   \quad\mu_\ast(\infty)=0.
\end{equation}     
Then (\ref{field2}) is straightforward to verify, and 
\begin{equation}\label{emuDGL} 
   \frac{d}{dr}\,e^{\mu_\ast(r)}
   =e^{\mu_\ast(r)+2\lambda_\ast(r)}\Big(\frac{m_\ast(r)}{r^2}+4\pi r\,p_\ast(r)\Big)
\end{equation} 
holds. 
Also 
\[ 0=F(f_\ast(r, \cdot, \cdot), r, \rho_\ast(r))
   =\int_0^\infty d\beta\int_{\R} dw\,
   \sqrt{1+w^2+\beta/r^2}\,f_\ast(r, w, \beta)
   -\frac{r^2}{\pi}\,\rho_\ast(r) \]   
by (\ref{F0}) and the definition of $F$. As a consequence, 
\[ \rho_{f_\ast}=\int dv\,\sqrt{1+v^2}\,f_\ast
   =\frac{\pi}{r^2}\int_0^\infty d\beta\int_{\R} dw
   \,\sqrt{1+w^2+\beta/r^2}\,f_\ast=\rho_\ast, \]    
and in particular $4\pi\int_0^\infty r^2\,\rho_{f_\ast}(r)\,dr=M$. 
Also $\rho_\ast(r)=0$ for $r\in [R_\ast, \infty[$ 
in conjunction with (\ref{iap}) and Lemma \ref{wibirk} yields 
$f_\ast(r, w, \beta)=0$ for $r\in [R_\ast, \infty[$. 
We also note that since $m_{\ast}(r)=M$, $p_{\ast}(r)=0$ 
and $\lambda_\ast(r)=-\frac{1}{2}\ln(1-\frac{2M}{r})$ for $r\geq R_\ast$, 
it follows from (\ref{muast_intro}) through integration that 
\begin{equation}\label{muR}
   -\mu_{\ast}(R_\ast)=M\int_{R_\ast}^\infty\frac{dr}{r^2-2Mr}
   =-\frac{1}{2}\,\ln\Big(1-\frac{2M}{R_\ast}\Big).
\end{equation}
\smallskip 

It remains to see that $f_\ast$ has the desired form, which is the main part 
of the argument. We start with a few observations, 
some of which are close to what has been attempted in \cite{Wol}. 

\begin{lemma}\label{abw1} For $r>0$, $\chi\in\Xi_r$, $a>0$ 
and $\eps\in [\frac{1}{\sqrt{2}}\,\alpha, \alpha]$ let
\begin{eqnarray}\label{kleinh}  
   h(\chi, r, a, \eps) & = & H(\chi)+\eps F(\chi, r, a)
   \nonumber
   \\ & = & \int_0^\infty d\beta\int_{\R} dw\,
   (\hat{\Phi}(\chi)-(\alpha-\eps\sqrt{1+w^2+\beta/r^2})\,\chi)
   -\eps\,\frac{r^2}{\pi}\,a
\end{eqnarray}  
and 
\[ \tilde{h}(r, a, \eps)=\inf_{\chi\in\Xi_r} h(\chi, r, a, \eps). \] 
Then 
\begin{equation}\label{fibi}  
   \tilde{h}(r, a, \eps)=-\Psi(\eps, r)-\eps\,\frac{r^2}{\pi}\,a 
\end{equation} 
at fixed $(r, a, \eps)$, and the infimum defining $\tilde{h}$ 
is attained at $\chi_\ast=\chi_\ast(\cdot, \cdot; r, a, \eps)\in\Xi_r$ given by 
\begin{equation}\label{chiast} 
   \chi_\ast(w, \beta; r, a, \eps)=(\alpha-\eps\sqrt{1+w^2+\beta/r^2})_+^\sigma
   =\phi(\alpha-\eps\sqrt{1+w^2+\beta/r^2}), 
\end{equation} 
i.e., we have 
\begin{equation}\label{ambru} 
   h(\chi_\ast(\cdot, \cdot; r, a, \eps), r, a, \eps)=\tilde{h}(r, a, \eps). 
\end{equation} 
\end{lemma} 
{\bf Proof\,:} To establish (\ref{fibi}), 
let first $\chi=\chi(w, \beta)\in\Xi_r$ be arbitrary. Then 
\[ (\alpha-\eps\sqrt{1+w^2+\beta/r^2})\,\chi(w, \beta)-\hat{\Phi}(\chi(w, \beta))
   \le\widehat{\widehat{\Phi}}(\alpha-\eps\sqrt{1+w^2+\beta/r^2}), \] 
and hence 
\[ h(\chi, r, a, \eps)\ge -\int_0^\infty d\beta\int_{\R} dw\,
   \widehat{\widehat{\Phi}}(\alpha-\eps\sqrt{1+w^2+\beta/r^2})-\eps\,\frac{r^2}{\pi}\,a, \] 
from where we deduce that also 
\[ \tilde{h}(r, a, \eps)\ge -\int_0^\infty d\beta\int_{\R} dw\,
   \widehat{\widehat{\Phi}}(\alpha-\eps\sqrt{1+w^2+\beta/r^2})-\eps\,\frac{r^2}{\pi}\,a. \] 
For the converse, for every $(w, \beta)$ consider the function 
\[ \varphi(s)=\hat{\Phi}(s)-(\alpha-\eps\sqrt{1+w^2+\beta/r^2})s,\quad s\in\R, \] 
and recall (\ref{ttl}). If $\alpha-\eps\sqrt{1+w^2+\beta/r^2}\le 0$, 
then the minimum of $\varphi$ is attained at $s_\ast=0$, where $\varphi(s_\ast)=0$. 
Secondly, if $\alpha-\eps\sqrt{1+w^2+\beta/r^2}\ge 0$, 
then the minimum of $\varphi$ is attained at $s_\ast=(\alpha-\eps\sqrt{1+w^2+\beta/r^2})^\sigma$, 
where $\varphi(s_\ast)=-\frac{1}{\sigma+1}\,(\alpha-\eps\sqrt{1+w^2+\beta/r^2})^{\sigma+1}$. 
Thus both cases can be summarized as follows: 
the minimum of $\varphi$ is attained at $s_\ast=(\alpha-\eps\sqrt{1+w^2+\beta/r^2})_+^\sigma$, 
where $\varphi(s_\ast)=-\frac{1}{\sigma+1}\,(\alpha-\eps\sqrt{1+w^2+\beta/r^2})_+^{\sigma+1}$. 
Let $\chi_\ast$ be defined as in (\ref{chiast}), and we drop the variables $(r, a, \eps)$ 
from its arguments to simplify notation. Then $\chi_\ast\in\Xi_r$, since $0\le\chi_\ast(w, \beta)
\le\alpha^\sigma$ shows that $\chi_\ast$ is bounded, and it has compact support, 
so $\chi_\ast\in L^{1+1/\sigma}(\R\times [0, \infty[)$. 
Furthermore, if $\sqrt{1+w^2+\beta/r^2}\ge 2$, then  
$\eps\sqrt{1+w^2+\beta/r^2}\ge 2\,\frac{1}{\sqrt{2}}\,\alpha>\alpha$ 
and hence $\chi_\ast(w, \beta)=0$. Therefore indeed $\chi_\ast\in\Xi_r$, 
and $s_\ast=\chi_\ast(w, \beta)$. Also 
\begin{eqnarray*} 
   \hat{\Phi}(\chi_\ast(w, \beta))-(\alpha-\eps\sqrt{1+w^2+\beta/r^2})
   \,\chi_\ast(w, \beta) 
   & = & \inf_{s\in\R}\,(\hat{\Phi}(s)-(\alpha-\eps\sqrt{1+w^2+\beta/r^2})s)
   \\ & = & -\sup_{s\in\R}\,(s(\alpha-\eps\sqrt{1+w^2+\beta/r^2})-\hat{\Phi}(s))
   \\ & = & -\,\widehat{\widehat{\Phi}}(\alpha-\eps\sqrt{1+w^2+\beta/r^2}). 
\end{eqnarray*} 
It follows that 
\begin{eqnarray*} 
   \tilde{h}(r, a, \eps)
   & \le & h(\chi_\ast, r, a, \eps)
   \\ & = & \int_0^\infty d\beta\int_{\R} dw\,(\hat{\Phi}(\chi_\ast)
   -(\alpha-\eps\sqrt{1+w^2+\beta/r^2})\,\chi_\ast)
   -\eps\,\frac{r^2}{\pi}\,a
   \\ & = & -\int_0^\infty d\beta\int_{\R} dw\,
   \widehat{\widehat{\Phi}}(\alpha-\eps\sqrt{1+w^2+\beta/r^2})
   -\eps\,\frac{r^2}{\pi}\,a,  
\end{eqnarray*} 
This finishes the proof of (\ref{fibi}), taking into account that 
$\widehat{\widehat{\Phi}}=\Phi$ and the definition of $\Psi$ in (\ref{somal}). 
The fact that the infimum is attained at $\chi_\ast$ 
is a consequence of the above argument, since this function 
realizes the pointwise minimum of the integrand 
$\hat{\Phi}(s)-(\alpha-\eps\sqrt{1+w^2+\beta/r^2})s$. 
{\hfill$\Box$}\bigskip  

\begin{lemma}\label{abw2} Let $r>0$ and $a\in [0, \eta]$. Then we have 
\[  \sup_{\eps\in [\frac{1}{\sqrt{2}}\,\alpha, \alpha]} \tilde{h}(r, a, \eps)
    =\sup_{\eps\in\R}\tilde{h}(r, a, \eps), \]
and the supremum is uniquely attained at $\eps_\ast=(G')^{-1}(-4\pi\kappa\,a)
\in [\frac{1}{\sqrt{2}}\,\alpha, \alpha]$, 
taking the value 
\begin{equation}\label{ameich} 
   \tilde{h}(r, a, \eps_\ast)
   =\sup_{\eps\in\R}\tilde{h}(r, a, \eps)
   =\hat{\Psi}\Big(-\frac{r^2}{\pi}\,a, r\Big)
   =\frac{4r^2}{\sigma+1}\,\hat{G}(-4\pi\kappa a).
\end{equation}  
\end{lemma} 
{\bf Proof\,:} From (\ref{fibi}) and (\ref{PsiG}) we obtain 
\[ \tilde{h}(r, a, \eps)=-\Psi(\eps, r)-\eps\,\frac{r^2}{\pi}\,a
   =-\frac{4r^2}{\sigma+1}\,\Big(G(\eps)+\eps\,\frac{\sigma+1}{4\pi}\,a\Big)
   =-\frac{4r^2}{\sigma+1}\,(G(\eps)+4\pi\kappa\,a\eps), \] 
and hence 
\[ \sup_{\eps\in I}\tilde{h}(r, a, \eps)
   =-\frac{4r^2}{\sigma+1}\,\inf_{\eps\in I}\,(G(\eps)+4\pi\kappa\,a\,\eps) \] 
for $I\subset\R$. Since $G(\eps)=\infty$ for $\eps\in ]-\infty, 0]$, 
no infimum is attained at such $\eps$. If $\eps\ge\alpha$, then $G(\eps)=0$, 
and hence $G(\eps)+4\pi\kappa\,a\,\eps\ge G(\alpha)+4\pi\kappa\,a\,\alpha$. 
For $\eps\in ]0, \alpha[$, the function $\varphi(\eps)=G(\eps)+4\pi\kappa\,a\,\eps$ 
has a minimum, where $\varphi'(\eps)=0$, which is at $\eps_\ast=(G')^{-1}(-4\pi\kappa a)$. 
The function $(G')^{-1}$ is increasing in $]-\infty, 0]$, 
and $0\le a\le\eta$, whence we deduce that $\eps_\ast\ge (G')^{-1}(-4\pi\kappa\eta)$. 
But ${\cal G}(s)=\hat{G}(-s)$ by definition, thus ${\cal G}'(s)=-\hat{G}'(-s)$ 
and (\ref{derLeg}) yields $\eps_\ast\ge (G')^{-1}(-4\pi\kappa\eta)=\hat{G}'(-4\pi\kappa\eta)
=-{\cal G}'(4\pi\kappa\eta)\ge\frac{1}{\sqrt{2}}\,\alpha$. 
This completes the proof of the lemma. 
{\hfill$\Box$}\bigskip  

\begin{lemma}\label{abw3} 
Let $r>0$, $a\in ]0, \eta[$ and $\eps_\ast=(G')^{-1}(-4\pi\kappa a)$. 
Then we have 
\[ F(\chi_\ast(\cdot, \cdot; r, a, \eps_\ast), r, a)=0. \]  
\end{lemma} 
{\bf Proof\,:} At $r>0$ and $a\in ]0, \eta[$ fixed 
we write $\chi_\ast(\eps)=\chi_\ast(\eps)(w, \beta)=\chi_\ast(w, \beta; r, a, \eps)$. 
Then, by Lemmas \ref{abw1} and \ref{abw2}, the function 
\[ \varphi(\eps)=h(\chi_\ast(\eps), r, a, \eps)=\tilde{h}(r, a, \eps)
   =-\frac{4r^2}{\sigma+1}\,(G(\eps)+4\pi\kappa\,a\eps) \] 
has a minimum at $\eps_\ast=(G')^{-1}(-4\pi\kappa a)
\in ]\frac{1}{\sqrt{2}}\,\alpha, \alpha[$, and thus $\varphi'(\eps_\ast)=0$. 
To calculate 
\[ \frac{d}{d\eps}\,h(\chi_\ast(\eps), r, a, \eps)\Big|_{\eps=\eps_\ast}, \] 
we can invoke (\ref{kleinh}), $\chi_\ast\ge 0$ and (\ref{ttl}) to obtain 
\begin{eqnarray*} 
   \varphi(\eps) & = & h(\chi, r, a, \eps)
   \\ & = & \int_0^\infty d\beta\int_{\R} dw\,
   \Big(\frac{\sigma}{\sigma+1}\,\chi_\ast(\eps)(w, \beta)^{1+1/\sigma}
   -(\alpha-\eps\sqrt{1+w^2+\beta/r^2})\,\chi_\ast(\eps)(w, \beta)\Big)
   -\eps\,\frac{r^2}{\pi}\,a. 
\end{eqnarray*} 
As the support of $\chi_\ast(\eps)$ is contained in 
$\{\alpha-\eps\sqrt{1+w^2+\beta/r^2}\ge 0\}$, 
we may rewrite this as 
\begin{eqnarray*} 
   \varphi(\eps) 
   & = & \int_0^\infty d\beta\int_{\R} dw\,{\bf 1}_{\{\alpha-\eps\sqrt{1+w^2+\beta/r^2}\ge 0\}}
   \Big(\frac{\sigma}{\sigma+1}\,\chi_\ast(\eps)(w, \beta)^{1+1/\sigma}
   \\ & & \hspace{15em} 
   -\,(\alpha-\eps\sqrt{1+w^2+\beta/r^2})\,\chi_\ast(\eps)(w, \beta)\Big)
   -\eps\,\frac{r^2}{\pi}\,a 
   \\ & = & 2r^2\int_0^\infty dy\,y\int_{\R} dw\,{\bf 1}_{\{\alpha-\eps\sqrt{1+w^2+y^2}\ge 0\}}
   \Big(\frac{\sigma}{\sigma+1}\,\chi_\ast(\eps)(w, r^2 y^2)^{1+1/\sigma}
   \\ & & \hspace{17em} 
   -\,(\alpha-\eps\sqrt{1+w^2+y^2})\,\chi_\ast(\eps)(w, r^2 y^2)\Big)
   -\eps\,\frac{r^2}{\pi}\,a  
   \\ & = & 2r^2\int_0^\pi d\theta\sin\theta
   \int_0^{\sqrt{\frac{\alpha^2}{\eps^2}-1}} d\xi\,\xi^2\,
   \Big(\frac{\sigma}{\sigma+1}\,\lambda_\ast(\eps)(\theta, \xi)^{1+1/\sigma}
   \\ & & \hspace{14em} 
   -\,(\alpha-\eps\sqrt{1+\xi^2})\,\lambda_\ast(\eps)(\theta, \xi)\Big)
   -\eps\,\frac{r^2}{\pi}\,a,   
\end{eqnarray*} 
where 
\[ \lambda_\ast(\eps)(\theta, \xi)
   =\chi_\ast(\eps)(\xi\cos\theta, r^2\xi^2\sin^2\theta)
   =(\alpha-\eps\sqrt{1+\xi^2})^\sigma \] 
by (\ref{chiast}) on the domain of integration, 
and we have used the changes of variables 
$y^2=\beta/r^2$, $2y\,dy=d\beta/r^2$ and thereafter 
$(w, y)=\xi(\cos\theta, \sin\theta)$, $|\det d(w, y)/d(\theta, \xi)|=\xi$. 
Noting that $\lambda_\ast(\eps)(\theta, \sqrt{\frac{\alpha^2}{\eps^2}-1})=0$, 
we can thus differentiate $\varphi$ close to $\eps_\ast$ to get  
\begin{eqnarray*} 
   \varphi'(\eps) 
   & = & 2r^2\int_0^\pi d\theta\sin\theta
   \int_0^{\sqrt{\frac{\alpha^2}{\eps^2}-1}} d\xi\,\xi^2\,
   \Big(\lambda_\ast(\eps)(\theta, \xi)^{1/\sigma}
   -(\alpha-\eps\sqrt{1+\xi^2})\Big)\lambda'_\ast(\eps)(\theta, \xi)
   \\ & & +\,2r^2\int_0^\pi d\theta\sin\theta
   \int_0^{\sqrt{\frac{\alpha^2}{\eps^2}-1}} d\xi\,\xi^2\,
   \sqrt{1+\xi^2}\,\lambda_\ast(\eps)(\theta, \xi)
   -\frac{r^2}{\pi}\,a.    
\end{eqnarray*}  
Since $\lambda_\ast(\eps)(\theta, \xi)^{1/\sigma}=\alpha-\eps\sqrt{1+\xi^2}$, 
this simplifies to 
\[ \varphi'(\eps) 
   =2r^2\int_0^\pi d\theta\sin\theta
   \int_0^{\sqrt{\frac{\alpha^2}{\eps^2}-1}} d\xi\,\xi^2\,
   \sqrt{1+\xi^2}\,\lambda_\ast(\eps)(\theta, \xi)
   -\frac{r^2}{\pi}\,a. \]
Thus if we undo the transformations, it is found that 
\begin{eqnarray*} 
   \varphi'(\eps) 
   & = & \int_0^\infty d\beta
   \int_{\R} dw\,{\bf 1}_{\{\alpha-\eps\sqrt{1+w^2+\beta/r^2}\ge 0\}}
   \sqrt{1+w^2+\beta/r^2}\,\chi_\ast(\eps)(w, \beta)
   -\frac{r^2}{\pi}\,a
   \\ & = & \int_0^\infty d\beta
   \int_{\R} dw\,\sqrt{1+w^2+\beta/r^2}\,\chi_\ast(\eps)(w, \beta)
   -\frac{r^2}{\pi}\,a
   \\ & = & F(\chi_\ast(\eps), r, a).  
\end{eqnarray*}    
As a consequence, 
\[ 0=\varphi'(\eps_\ast)=F(\chi_\ast(\eps_\ast), r, a), \] 
as was to be shown. 
{\hfill$\Box$}\bigskip 

\begin{cor}\label{mbcor} 
Let $r>0$, $a\in ]0, \eta[$ and $\eps_\ast=(G')^{-1}(-4\pi\kappa a)$. 
Then we have
\[ \chi_\ast(\cdot, \cdot; r, a, \eps_\ast)=\chi(\cdot, \cdot; r, a). \]   
In particular, 
\begin{equation}\label{fastform} 
   f_\ast(r, w, \beta)
   =\phi(\alpha-\tilde{\eps}(r)\sqrt{1+w^2+\beta/r^2}) 
\end{equation}
for $\tilde{\eps}(r)=(G')^{-1}(-4\pi\kappa\rho_\ast(r))$. 
\end{cor} 
{\bf Proof\,:} We continue to denote $\chi_\ast(\eps_\ast)
=\chi_\ast(\eps_\ast)(w, \beta)=\chi_\ast(w, \beta; r, a, \eps_\ast)$. 
Then $\chi_\ast(\eps_\ast)\in\Xi_r$ by Lemma \ref{abw1} and $F(\chi_\ast(\eps_\ast), r, a)=0$ 
due to Lemma \ref{abw3}. Furthermore, by the various definitions, 
\begin{eqnarray*} 
   H(\chi_\ast(\eps_\ast)) & = & H(\chi_\ast(\eps_\ast))+\eps_\ast F(\chi_\ast(\eps_\ast), r, a)
   =h(\chi_\ast(\eps_\ast), r, a, \eps_\ast)
   =\tilde{h}(r, a, \eps_\ast)
   \\ & = & \inf_{\chi\in\Xi_r} h(\chi, r, a,\eps_\ast)
   \le h(\chi(\cdot, \cdot; r, a), r, a, \eps_\ast)
   \\ & = & H(\chi(\cdot, \cdot; r, a))+\eps_\ast F(\chi(\cdot, \cdot; r, a), r, a)
   =H(\chi(\cdot, \cdot; r, a)). 
\end{eqnarray*} 
From the uniqueness of the minimizer (Lemma \ref{wibirk}) 
we thus deduce that $\chi_\ast(\eps_\ast)=\chi(\cdot, \cdot; r, a)$. 
For (\ref{fastform}) it suffices to recall (\ref{iap}) and 
to notice that here we take $a=\rho_\ast(r)$, 
$\rho_\ast\le\eta$ is strictly decreasing on $[0, R_\ast]$ 
and such that $\rho_\ast(R_\ast)=0$ by Proposition \ref{EL_exi}, 
whence $\rho_\ast(r)\in ]0, \eta[$ for $r\in ]0, R_\ast[$. 
{\hfill$\Box$}\bigskip  

\begin{cor}\label{wolrel} Let $r>0$ and $a\in ]0, \eta[$. Then 
\[ \hat{\Psi}\Big(-\frac{r^2}{\pi}\,a, r\Big)=H(\chi(\cdot, \cdot; r, a)). \]  
\end{cor} 
{\bf Proof\,:} Define $\eps_\ast=(G')^{-1}(-4\pi\kappa a)$ as before. 
Then $\chi_\ast(\cdot, \cdot; r, a, \eps_\ast)=\chi(\cdot, \cdot; r, a)$ 
by Corollary \ref{mbcor}, and furthermore $F(\chi(\cdot, \cdot; r, a), r, a)
=F(\chi_\ast(\cdot, \cdot; r, a, \eps_\ast), r, a)=0$ by Lemma \ref{abw3}. 
Therefore (\ref{ameich}) and (\ref{ambru}) yield
\begin{eqnarray*} 
   \hat{\Psi}\Big(-\frac{r^2}{\pi}\,a, r\Big) 
   & = & \tilde{h}(r, a, \eps_\ast)=h(\chi_\ast(\cdot, \cdot; r, a, \eps_\ast), r, a, \eps_\ast)
   =h(\chi(\cdot, \cdot; r, a), r, a, \eps_\ast)
   \\ & = & H(\chi(\cdot, \cdot; r, a))+\eps_\ast F(\chi(\cdot, \cdot; r, a), r, a)
   =H(\chi(\cdot, \cdot; r, a)), 
\end{eqnarray*} 
as claimed. 
{\hfill$\Box$}\bigskip 

We still need to verify that the argument on the right-hand side 
of (\ref{fastform}) is a function of the energy 
$E_\ast=e^{\mu_\ast(r)}\sqrt{1+v^2}=e^{\mu_\ast(r)}\sqrt{1+w^2+\beta/r^2}$. 
For this we have to relate $\tilde{\eps}(r)$ 
to $e^{\mu_\ast(r)}$, and here the Euler-Lagrange equation (\ref{ELmast}) for $m_\ast$ 
enters in a crucial way; we will use it in the form $l(r)=0$ for $r\in [0, R_\ast]$, 
cf.~Remark \ref{kleinl_intro}. Hence it follows from (\ref{kleinl_form}) and (\ref{PsiG}) that 
\begin{eqnarray}\label{phmo}  
   0 & = & -\zeta'(r)+\frac{m_\ast(r)}{r^2}\,\frac{1}{1-\frac{2m_\ast(r)}{r}}\,\zeta(r)
   -\frac{1}{\kappa}\,\frac{r}{1-\frac{2m_\ast(r)}{r}}\,G(-\zeta(r))
   \nonumber
   \\ & = & -\zeta'(r)+\frac{m_\ast(r)}{r^2}\,\frac{1}{1-\frac{2m_\ast(r)}{r}}\,\zeta(r)
   -\frac{4\pi^2}{r^2}\,\frac{r}{1-\frac{2m_\ast(r)}{r}}
   \,\Psi(-\zeta(r), r)
\end{eqnarray}
for $\zeta(r)={\cal G}'(\kappa\frac{m'_\ast(r)}{r^2})
={\cal G}'(4\pi\kappa\rho_\ast(r))=-\hat{G}'(-4\pi\kappa\rho_\ast(r))$. 
From (\ref{derLeg}) we deduce that   
\begin{equation}\label{bsit} 
   \tilde{\eps}(r)=(G')^{-1}(-4\pi\kappa\rho_\ast(r))
   =\hat{G}'(-4\pi\kappa\rho_\ast(r)),
\end{equation}    
which shows that $\zeta(r)=-\tilde{\eps}(r)$. 
Therefore (\ref{phmo}) comes down to 
\[ 0=\tilde{\eps}'(r)
   -\frac{1}{r-2m_\ast(r)}\,\frac{m_\ast(r)}{r}\,\tilde{\eps}(r)
   -\frac{4\pi^2}{r-2m_\ast(r)}\,\Psi(\tilde{\eps}(r), r). \] 
Since $e^{2\lambda_\ast(r)}=\frac{r}{r-2m_\ast(r)}$, we get  
\begin{equation}\label{schmi} 
   \tilde{\eps}'(r)
   =e^{2\lambda_\ast(r)}\,\Big(\frac{m_\ast(r)}{r^2}\,\tilde{\eps}(r)
   +\frac{4\pi^2}{r}\,\Psi(\tilde{\eps}(r), r)\Big).
\end{equation} 
Lastly, from (\ref{Psi2}), (\ref{fastform}) and (\ref{pastform}) we obtain 
\begin{eqnarray*} 
   \Psi(\tilde{\eps}(r), r)
   & = & \tilde{\eps}(r)\int_0^\infty d\beta\int_{\R} dw\,\frac{w^2}{\sqrt{1+w^2+\beta/r^2}}
   \,\phi(\alpha-\tilde{\eps}(r)\sqrt{1+w^2+\beta/r^2})
   \\ & = & \tilde{\eps}(r)\int_0^\infty d\beta\int_{\R} dw\,\frac{w^2}{\sqrt{1+w^2+\beta/r^2}}
   \,f_\ast(r, w, \beta)
   \\ & = & \frac{r^2}{\pi}\,\tilde{\eps}(r)\,p_\ast(r). 
\end{eqnarray*} 
Hence (\ref{schmi}) shows that 
\[ \tilde{\eps}'(r)
   =\tilde{\eps}(r)\,e^{2\lambda_\ast(r)}\,\Big(\frac{m_\ast(r)}{r^2}+4\pi r\,p_\ast(r)\Big). \] 
Comparing this to (\ref{emuDGL}), it follows that 
\[ \frac{d}{dr}\,(e^{-\mu_\ast(r)}\tilde{\eps}(r))=0,\,\,\, 0\leq r \leq R_\ast, \] 
so that $e^{-\mu_\ast(r)}\tilde{\eps}(r)=c_\ast$ for some constant $c_\ast$. 
Thus $\tilde{\eps}(r)=c_\ast e^{\mu_\ast(r)}$. Next we have from (\ref{muR})
that $e^{\mu_\ast(R_\ast)}=\sqrt{1-2M/R_\ast}$. 
Furthermore, $\tilde{\eps}(r)=\hat{G}'(-4\pi\kappa\rho_\ast(r))$ 
by (\ref{bsit}) together with $\rho_\ast(R_\ast)=0$ and $\hat{G}'(0)=\alpha$ 
leads to $\tilde{\eps}(R_\ast)=\alpha$, which in turn yields 
\[ c_\ast=\frac{\alpha}{\sqrt{1-\frac{2M}{R_\ast}}}. \] 
Therefore (\ref{fastform}) finally implies that 
\[ f_\ast(r, w, \beta)
   =\phi\Big(\alpha-\frac{\alpha}{\sqrt{1-\frac{2M}{R_\ast}}}\,e^{\mu_\ast(r)}\sqrt{1+w^2+\beta/r^2}\Big), \] 
which has the desired form, depending only on $E_\ast$. 
This completes the proof of Theorem \ref{mainthm1}. 
{\hfill$\Box$}\bigskip


\section{Relations to stability}
\label{rem_sect}   
\setcounter{equation}{0} 

Consider the particle number-Casimir functional ${\cal D}$ from (\ref{pnC}). 
We begin with a lemma that makes a relation between ${\cal D}$ 
and ${\cal L}$, see (\ref{holte}). 

\begin{lemma}\label{rowi} 
Let $\alpha>0$ be fixed and suppose that $\eta>0$ 
is such that (\ref{etacond}) is verified. 
Let a static solution $(f_\ast, \lambda_\ast, \mu_\ast)$ 
be constructed as in Theorem \ref{mainthm1}. Then 
\begin{itemize} 
\item[(a)] ${\cal L}(m_\ast)=\kappa\,{\cal D}(f_\ast)$; 
\item[(b)] if $(f, \lambda, \mu)$ is a further (possibly time-dependent) solution so that  
\[ f(t, r, \cdot, \cdot)\in\Xi_r,\quad\frac{2m(t, r)}{r}<1,\quad\rho(t, r)\in ]0, \eta[, \] 
for $t\in [0, T]$ and $r\in ]0, \infty[$, then ${\cal L}(m(t))\le\kappa\,{\cal D}(f(t))$ 
for $t\in [0, T]$, where $m(t)(r)=m(t, r)$ and $f(t)(r, w, \beta)=f(t, r, w, \beta)$. 
\end{itemize} 
\end{lemma} 
{\bf Proof\,:} (a) We have $\rho_{f_\ast}=\rho_\ast$, $\lambda_{f_\ast}=\lambda_\ast$ 
and $e^{-2\lambda_\ast(r)}=1-\frac{2m_\ast(r)}{r}$. 
Hence by (\ref{xvtransf}), (\ref{HF-def}), and due to $f_\ast(r, w, \beta)=0$ for $r\in [R_\ast, \infty[$: 
\begin{eqnarray*} 
   {\cal D}(f_\ast) & = & \int_{\R^3}\int_{\R^3} e^{\lambda_\ast}\,(\hat{\Phi}(f_\ast)-\alpha f_\ast)\,dx\,dv
   \\[1ex] & = & 4\pi^2\int_0^\infty dr\,\sqrt{\frac{r}{r-2m_\ast(r)}}\int_0^\infty d\beta\int_{\R} dw\,
   (\hat{\Phi}(f_\ast)-\alpha f_\ast)
   \\[1ex] & = & 4\pi^2\int_0^{R_\ast} dr\,\sqrt{\frac{r}{r-2m_\ast(r)}}\int_0^\infty d\beta\int_{\R} dw\,
   (\hat{\Phi}(f_\ast)-\alpha f_\ast)
   \\[1ex] & = & 4\pi^2\int_0^{R_\ast}\sqrt{\frac{r}{r-2m_\ast(r)}}
   \,H(f_\ast(r, \cdot, \cdot))\,dr
   \\[1ex] & = & 4\pi^2\int_0^{R_\ast}\sqrt{\frac{r}{r-2m_\ast(r)}}
   \,H(\chi(\cdot, \cdot; r, \rho_\ast(r)))\,dr. 
\end{eqnarray*} 
Also $\frac{m'_\ast(r)}{r^2}=4\pi\rho_\ast(r)=0$ for $r\in [R_\ast, \infty[$. 
Since $\hat{G}(0)={\cal G}(0)=0$ by Lemma \ref{calGprops}, it follows that 
\begin{eqnarray*}
   {\cal L}(m_\ast) & = & \int_0^\infty L(r, m_\ast(r), m'_\ast(r))\,dr
   \\[1ex] & = & \int_0^{R_\ast}\frac{r^{5/2}}{\sqrt{r-2m_\ast(r)}}
   \,\hat{G}\Big(-\kappa\,\frac{m'_\ast(r)}{r^2}\Big)\,dr
   \\[1ex] & = & \frac{\sigma+1}{4}\int_0^{R_\ast}\sqrt{\frac{r}{r-2m_\ast(r)}}
   \,\hat{\Psi}\Big(-\frac{r^2}{\pi}\,\rho_\ast(r), r\Big)\,dr, 
\end{eqnarray*}
where we have used (\ref{rati2}) in the last step. 
It remains to apply Corollary \ref{wolrel} for $a=\rho_\ast(r)$. 
(b) In the same way as in (a) we obtain 
\begin{eqnarray*}
   {\cal D}(f(t)) & = & 4\pi^2\int_0^\infty\sqrt{\frac{r}{r-2m(r)}}
   \,H(f(t, r, \cdot, \cdot))\,dr, 
   \\[1ex] 
   {\cal L}(m(t)) & = & \frac{\sigma+1}{4}\int_0^\infty\sqrt{\frac{r}{r-2m(t, r)}}
   \,\hat{\Psi}\Big(-\frac{r^2}{\pi}\,\rho(t, r), r\Big)\,dr. 
\end{eqnarray*} 
Since $\rho(t, r)\in ]0, \eta[$, Corollary \ref{wolrel} implies that 
$\hat{\Psi}(-\frac{r^2}{\pi}\,\rho(t, r), r)=H(\chi(\cdot, \cdot; r, \rho(t, r)))$. 
By (\ref{rho2}) we have 
\[ \frac{r^2}{\pi}\,\rho(t, r)=\int_{\R}\int_0^\infty\sqrt{1+w^2+\beta/r^2}
   \,f(t, r, w, \beta)\,d\beta\,dw, \] 
which means that $F(f(t, r, \cdot, \cdot), r, \rho(r))=0$. 
As we are assuming that $f(t, r, \cdot, \cdot)\in\Xi_r$, 
(\ref{lslem}) shows that $H(\chi(\cdot, \cdot; r, \rho(t, r)))
\le H(f(t, r, \cdot, \cdot))$. Altogether, this yields 
${\cal L}(m(t))\le\frac{\sigma+1}{16\pi^2}\,{\cal D}(f(t))=\kappa\,{\cal D}(f(t))$ 
for $t\in [0, T]$. 
{\hfill$\Box$}\bigskip

Note that since ${\cal D}$ is constant along solutions, in fact ${\cal D}(f(t))
={\cal D}(f(0))$ for $t\in [0, T]$ in part (b). We would like to be able to say 
that $f_\ast$ minimizes ${\cal D}$ over a certain class of functions ${\cal F}$, 
\begin{equation}\label{glme} 
   {\cal D}(f)\ge {\cal D}(f_\ast),\quad f\in {\cal F}.
\end{equation} 
In view of Lemma \ref{rowi}, for this we could resort to a general method, 
as for instance outlined in \cite[Prop.~1.18]{BGH}. We write 
\[ {\cal D}(f)-{\cal D}(f_\ast)=[{\cal D}(f)-\kappa^{-1}{\cal L}(m)]
   +[\kappa^{-1}{\cal L}(m)-\kappa^{-1}{\cal L}(m_\ast)] \] 
and we already know that the first $[\ldots]$ is non-negative. 
Thus to establish (\ref{glme}), the real issue is to show that 
$m_\ast$ minimizes ${\cal L}$ over a certain set of functions 
related to ${\cal F}$. In other words, since $m_\ast$ solves 
the associated Euler-Lagrange equation, it has to be clarified 
if this in turn does imply that $m_\ast$ is a minimizer 
(or if some condition has to be added). Generally speaking, 
for some variational problems this is possible, 
using ``Mayer fields''; see \cite{BGH} or \cite[Thm.~4.18]{Daco}. 


\section{Appendix}
\label{app_sect}   
\setcounter{equation}{0}  

\subsection{Properties of $G$, $G'$, $G''$ and ${\cal G}$}
\label{gs_sect} 

From (\ref{Gdef}) recall that 
\[ G(\eps)=\int_0^\infty\xi^2
   \,{(\alpha-\eps\sqrt{1+\xi^2})}_+^{\sigma+1}\,d\xi \] 
for $\sigma\in ]0, \frac{3}{2}[$, where it is understood that $G(\eps)=\infty$ 
for $\eps\in ]-\infty, 0]$. Here $\alpha>0$ is a parameter. 
Clearly $G(\eps)=0$ for $\eps\in [\alpha, \infty[$. 
\medskip 

Henceforth we are going to write $A\sim B$ for two functions 
$A(x)\ge 0$ and $B(x)\ge 0$, if there are constants $c_1, c_2>0$ 
such that $c_1 A(x)\le B(x)\le c_2 A(x)$ for all $x$, 
where $A$ and $B$ are defined. In other words, 
in each quantitative estimate (from above or from below) 
$A$ could be exchanged by $B$, and vice versa. 

\begin{lemma}\label{Philem} For $a\ge 0$ and $b>-1$ let 
\begin{equation}\label{Phiab} 
   \Phi_{a, b}(\theta)=\int_0^\theta\xi^2({\sqrt{1+\xi^2})}^a\,
   \frac{{(\theta^2-\xi^2)}^b}{{(\sqrt{1+\theta^2}+\sqrt{1+\xi^2})}^b}\,d\xi,
   \quad\theta\in [0, \infty[.
\end{equation} 
Then $\Phi_{a, b}(\theta)\sim\theta^{3+2b}$ for $\theta\in [0, 1]$ 
and $\Phi_{a, b}(\theta)\sim\theta^{3+b+a}$ for $\theta\in [1, \infty[$. 
\end{lemma} 
{\bf Proof\,:} To begin with, if $0\le\xi\le\theta$, then 
\[ \sqrt{1+\theta^2}\le\sqrt{1+\theta^2}+\sqrt{1+\xi^2}\le 2\sqrt{1+\theta^2}, \]
so that 
\[ \Phi_{a, b}(\theta)\sim\frac{1}{{(1+\theta^2)}^{b/2}}
   \int_0^\theta\xi^2({\sqrt{1+\xi^2})}^a\,{(\theta^2-\xi^2)}^b\,d\xi. \] 
Also $\theta\le\theta+\xi\le 2\theta$, which results in 
\[ \Phi_{a, b}(\theta)\sim\frac{\theta^b}{{(1+\theta^2)}^{b/2}}
   \int_0^\theta\xi^2({\sqrt{1+\xi^2})}^a {(\theta-\xi)}^b\,d\xi
   =\frac{\theta^{3+2b}}{{(1+\theta^2)}^{b/2}}
   \int_0^1 s^2({\sqrt{1+\theta^2 s^2})}^a {(1-s)}^b\,ds, \]   
where we have changed variables as $\xi=\theta s$, $d\xi=\theta\,ds$, 
in the second step; note that the integral is non-singular due to $b>-1$. 
If $\theta\in [0, 1]$, then $\sqrt{1+\theta^2 s^2}\sim 1$ for $s\in [0, 1]$ 
and $1+\theta^2\sim 1$. It follows that $\Phi_{a, b}(\theta)\sim\theta^{3+2b}$ 
in this case. On the other hand, if $\theta\in [1, \infty[$, 
then $1+\theta^2\sim\theta^2$ and moreover $\theta s\le\sqrt{1+\theta^2 s^2}\le\sqrt{2}\,\theta$ 
for $s\in [0, 1]$ yields $\Phi_{a, b}(\theta)\sim\theta^{3+b+a}$. 
{\hfill$\Box$}\bigskip 

\begin{cor}\label{PhiLip} 
Let $a\ge 0$ and $b>-1$. There is a constant $C_\ast>0$ such that 
\[ 	|\Phi_{a, b}(\theta)-\Phi_{a, b}(\tilde{\theta})|\le C_\ast |\theta-\tilde{\theta}|,
    \quad\theta, \tilde{\theta}\in [0, 1]. \] 
\end{cor} 
{\bf Proof\,:} In the case where $b\ge 0$ we may simply differentiate $\Phi_{a, b}(\theta)$ 
and bound the derivative. Thus we may assume that $b\in ]-1, 0[$ to rewrite $\Phi_{a, b}(\theta)$ as 
\begin{eqnarray}\label{sdn} 
   \Phi_{a, b}(\theta) 
   & = & \int_0^\theta\xi^2({\sqrt{1+\xi^2})}^a\,
   \frac{{(\sqrt{1+\theta^2}+\sqrt{1+\xi^2})}^{\hat{b}}}{{(\theta^2-\xi^2)}^{\hat{b}}}\,d\xi
   \nonumber
   \\ & = & \theta^{3-2\hat{b}}\int_0^1\tau^2 ({\sqrt{1+\theta^2\tau^2})}^a\,
   \frac{{(\sqrt{1+\theta^2}+\sqrt{1+\theta^2\tau^2})}^{\hat{b}}}{{(1-\tau^2)}^{\hat{b}}}\,d\tau
\end{eqnarray} 
for $\hat{b}=-b\in ]0, 1[$, and we used the change of variables $\xi=\theta\tau$, $d\xi=\theta d\tau$; 
note that $3-2\hat{b}\in ]1, 3[$. Since 
\[ \int_0^1\frac{1}{{(1-\tau)}^{\hat{b}}}\,d\tau<\infty \] 
is integrable, we can differentiate (\ref{sdn}) to obtain the expression
\begin{eqnarray*}
   \Phi_{a, b}(\theta) 
   & = & (3-2\hat{b})\,\theta^{2(1-\hat{b})}\int_0^1\tau^2 ({\sqrt{1+\theta^2\tau^2})}^a\,
   \frac{{(\sqrt{1+\theta^2}+\sqrt{1+\theta^2\tau^2})}^{\hat{b}}}{{(1-\tau^2)}^{\hat{b}}}\,d\tau
   \\ & & +\,a\,\theta^{4-2\hat{b}}\int_0^1\tau^4 ({\sqrt{1+\theta^2\tau^2})}^{a-2}\, 
   \frac{{(\sqrt{1+\theta^2}+\sqrt{1+\theta^2\tau^2})}^{\hat{b}}}{{(1-\tau^2)}^{\hat{b}}}\,d\tau
   \\ & & +\,\hat{b}\,\theta^{4-2\hat{b}}\int_0^1\tau^2 ({\sqrt{1+\theta^2\tau^2})}^a\,
   \frac{1}{{(\sqrt{1+\theta^2}+\sqrt{1+\theta^2\tau^2})}^{1-\hat{b}}{(1-\tau^2)}^{\hat{b}}}
   \\ & & \hspace{6em}\times\,\Big(\frac{1}{\sqrt{1+\theta^2}}+\frac{\tau^2}{\sqrt{1+\theta^2\tau^2}}\Big)\,d\tau, 
\end{eqnarray*} 
which is bounded in $\theta\in [0, 1]$. 
{\hfill$\Box$}\bigskip

\begin{lemma}[Properties of $G$]\label{Geasy} We have 
\begin{equation}\label{abm} 
   G(\eps)\sim\left\{\begin{array}{c@{\quad:\quad}l}
   \alpha^{-3/2}\,(\alpha-\eps)^{5/2+\sigma}  & \eps\in [\frac{1}{\sqrt{2}}\,\alpha, \alpha] \\[1ex] 
   \eps^{-3}\,\alpha^{4+\sigma} & \eps\in ]0, \frac{1}{\sqrt{2}}\,\alpha] 
   \end{array}\right. . 
\end{equation} 
In particular, since $G(\eps)=0$ for $\eps\in [\alpha, \infty[$, 
$G: ]0, \infty[\to\R$ is continuous. Also $G(\eps)>0$ for $\eps\in ]0, \alpha[$. 
\end{lemma} 
{\bf Proof\,:} Denoting $\tau=\alpha/\eps$ and $\theta=\sqrt{\tau^2-1}$, 
we see that for $\eps\in ]0, \alpha]$, and thus $\tau\in [1, \infty[$ 
as well as $\theta\in [0, \infty[$, 
\begin{eqnarray*} 
   G(\eps) & = & \int_0^\infty\xi^2\,{(\alpha-\eps\sqrt{1+\xi^2})}_+^{\sigma+1}\,d\xi
   =\eps^{\sigma+1}\int_0^\infty\xi^2\,{(\tau-\sqrt{1+\xi^2})}_+^{\sigma+1}\,d\xi
   \\ & = & \eps^{\sigma+1}\int_0^{\sqrt{\tau^2-1}}\xi^2\,{(\tau-\sqrt{1+\xi^2})}^{\sigma+1}\,d\xi
   =\eps^{\sigma+1}\int_0^{\sqrt{\tau^2-1}}\xi^2\,
   \frac{{(\tau^2-1-\xi^2)}^{\sigma+1}}
   {{(\tau+\sqrt{1+\xi^2})}^{\sigma+1}}\,d\xi
   \\ & = & \eps^{\sigma+1}\int_0^\theta\xi^2\,
   \frac{{(\theta^2-\xi^2)}^{\sigma+1}}
   {{(\sqrt{1+\theta^2}+\sqrt{1+\xi^2})}^{\sigma+1}}\,d\xi
   =\eps^{\sigma+1}\Phi_{0, \sigma+1}(\theta), 
\end{eqnarray*}    
cf.~(\ref{Phiab}). Therefore Lemma \ref{Philem} implies that 
\begin{eqnarray*} 
   G(\eps) & \sim & \eps^{\sigma+1}\,\left\{\begin{array}{c@{\quad:\quad}l}
   \theta^{5+2\sigma} & \theta\in [0, 1] \\[1ex] 
   \theta^{4+\sigma}  & \theta\in [1, \infty[\end{array}\right.
   \\[1ex] & = & \eps^{\sigma+1}\,\left\{\begin{array}{c@{\quad:\quad}l}
   (\tau^2-1)^{5/2+\sigma}  & \tau\in [1, \sqrt{2}] \\[1ex] 
   (\tau^2-1)^{2+\sigma/2}  & \tau\in [\sqrt{2}, \infty[
   \end{array}\right.
   \\[1ex] & = & \left\{\begin{array}{c@{\quad:\quad}l}
   \eps^{-4-\sigma}\,(\alpha^2-\eps^2)^{5/2+\sigma}  & \eps\in [\frac{1}{\sqrt{2}}\,\alpha, \alpha] \\[1ex] 
   \eps^{-3}\,(\alpha^2-\eps^2)^{2+\sigma/2}  & \eps\in ]0, \frac{1}{\sqrt{2}}\,\alpha] 
   \end{array}\right. . 
\end{eqnarray*} 
It remains to make use of the facts that $\alpha+\eps\sim\alpha$ for $\eps\in [0, \alpha]$, 
$\eps\sim\alpha$ for $\eps\in [\frac{1}{\sqrt{2}}\,\alpha, \alpha]$ 
and $\alpha-\eps\sim\alpha$ for $\eps\in ]0, \frac{1}{\sqrt{2}}\,\alpha]$. 
{\hfill$\Box$}\bigskip 

\begin{lemma}[Properties of $G'$]\label{Gesasy} We have 
\begin{equation}\label{Gsdef} 
   G'(\eps)=-(\sigma+1)\int_0^\infty
   \xi^2\sqrt{1+\xi^2}\,{(\alpha-\eps\sqrt{1+\xi^2})}_+^\sigma\,d\xi,
   \quad\eps\in ]0, \alpha],
\end{equation} 
and 
\begin{equation}\label{adv} 
   G'(\eps)\sim\left\{\begin{array}{l@{\quad:\quad}l}
   -\,\alpha^{-3/2}\,(\alpha-\eps)^{3/2+\sigma}  & \eps\in [\frac{1}{\sqrt{2}}\,\alpha, \alpha] \\[1ex] 
   -\,\eps^{-4}\,\alpha^{4+\sigma} & \eps\in ]0, \frac{1}{\sqrt{2}}\,\alpha] 
   \end{array}\right. . 
\end{equation} 
In particular, since $G'(\eps)=0$ for $\eps\in ]\alpha, \infty[$, 
$G': ]0, \infty[\to\R$ is continuous. Also $G'(\eps)<0$ for $\eps\in ]0, \alpha[$.   
\end{lemma} 
{\bf Proof\,:} By definition we may write 
\begin{equation}\label{yaG} 
   G(\eps)=\int_0^{\sqrt{\frac{\alpha^2}{\eps^2}-1}}
   \xi^2\,{(\alpha-\eps\sqrt{1+\xi^2})}^{\sigma+1}\,d\xi.
\end{equation} 
Differentiating w.r.~to $\eps$, the upper integration limit term 
makes no contribution, since $(\ldots)^{\sigma+1}$ vanishes at this point $\xi$; 
thus (\ref{Gsdef}) follows. Concerning the asymptotics of $G'(\eps)$, 
we can follow the same steps as in the proof of Lemma \ref{Geasy} to obtain the relation 
\[ G'(\eps)=-(\sigma+1)\,\eps^\sigma\,\Phi_{1, \sigma}(\theta). \] 
Therefore we can invoke Lemma \ref{Philem} to get 
\begin{eqnarray*} 
   G'(\eps) & \sim & -\,\eps^\sigma\,\left\{\begin{array}{c@{\quad:\quad}l}
   \theta^{3+2\sigma} & \theta\in [0, 1] \\[1ex] 
   \theta^{4+\sigma}  & \theta\in [1, \infty[\end{array}\right.
   \\[1ex] & = & -\,\eps^\sigma\,\left\{\begin{array}{c@{\quad:\quad}l}
   (\tau^2-1)^{3/2+\sigma}  & \tau\in [1, \sqrt{2}] \\[1ex] 
   (\tau^2-1)^{2+\sigma/2}  & \tau\in [\sqrt{2}, \infty[
   \end{array}\right.
   \\[1ex] & = & \left\{\begin{array}{l@{\quad:\quad}l}
   -\,\eps^{-3-\sigma}\,(\alpha^2-\eps^2)^{3/2+\sigma}  
   & \eps\in [\frac{1}{\sqrt{2}}\,\alpha, \alpha] \\[1ex] 
   -\,\eps^{-4}\,(\alpha^2-\eps^2)^{2+\sigma/2}  & \eps\in ]0, \frac{1}{\sqrt{2}}\,\alpha] 
   \end{array}\right. , 
\end{eqnarray*}  
which in turn leads to (\ref{adv}).  
{\hfill$\Box$}\bigskip 

\begin{lemma}[Properties of $G''$]\label{Gessasy} We have 
\begin{equation}\label{Gssdef} 
   G''(\eps)=\sigma (\sigma+1)\int_0^\infty
   \xi^2 (1+\xi^2)\,{(\alpha-\eps\sqrt{1+\xi^2})}_+^{\sigma-1}\,d\xi,
   \quad\eps\in ]0, \alpha],
\end{equation} 
and 
\begin{equation}\label{tawe} 
   G''(\eps)\sim\left\{\begin{array}{l@{\quad:\quad}l}
   \alpha^{-3/2}\,(\alpha-\eps)^{1/2+\sigma}  & \eps\in [\frac{1}{\sqrt{2}}\,\alpha, \alpha] \\[1ex] 
   \eps^{-5}\,\alpha^{4+\sigma} & \eps\in ]0, \frac{1}{\sqrt{2}}\,\alpha] 
   \end{array}\right. . 
\end{equation} 
In particular, since $G''(\eps)=0$ for $\eps\in ]\alpha, \infty[$, 
$G'': ]0, \infty[\to\R$ is continuous. Also $G''(\eps)>0$ for $\eps\in ]0, \alpha[$.    
\end{lemma} 
{\bf Proof\,:} Relation (\ref{Gssdef}) is derived from (\ref{Gsdef}) 
in the same way as (\ref{Gsdef}) is obtained from (\ref{yaG}); for this it is important 
that $\sigma>0$. Also we may rewrite $G''(\eps)$ in the form 
\begin{equation}\label{madu} 
   G''(\eps)=\sigma (\sigma+1)\,\eps^{\sigma-1}\,\Phi_{2, \sigma-1}(\theta).
\end{equation}  
Hence Lemma \ref{Philem} implies that  
\begin{eqnarray*} 
   G''(\eps) & \sim & \eps^{\sigma-1}\,\left\{\begin{array}{c@{\quad:\quad}l}
   \theta^{1+2\sigma} & \theta\in [0, 1] \\[1ex] 
   \theta^{4+\sigma}  & \theta\in [1, \infty[\end{array}\right.
   \\[1ex] & = & \eps^{\sigma-1}\,\left\{\begin{array}{c@{\quad:\quad}l}
   (\tau^2-1)^{1/2+\sigma}  & \tau\in [1, \sqrt{2}] \\[1ex] 
   (\tau^2-1)^{2+\sigma/2}  & \tau\in [\sqrt{2}, \infty[
   \end{array}\right.
   \\[1ex] & = & \left\{\begin{array}{l@{\quad:\quad}l}
   \eps^{-2-\sigma}\,(\alpha^2-\eps^2)^{1/2+\sigma}  & \eps\in [\frac{1}{\sqrt{2}}\,\alpha, \alpha] \\[1ex] 
   \eps^{-5}\,(\alpha^2-\eps^2)^{2+\sigma/2}  & \eps\in ]0, \frac{1}{\sqrt{2}}\,\alpha] 
   \end{array}\right. , 
\end{eqnarray*}
and this gives (\ref{tawe}). 
{\hfill$\Box$}\bigskip 

\begin{cor}\label{GessLip} 
For every $q\in [\frac{1}{\sqrt{2}}, 1[$ there exists a constant $C_q>0$ 
such that 
\[ |G''(\eps)-G''(\tilde{\eps})|\le C_q\,\alpha^{-(2-\sigma)}\,|\eps-\tilde{\eps}|,
   \quad\eps, \tilde{\eps}\in \Big[\frac{1}{\sqrt{2}}\,\alpha, q\alpha\Big]. \] 
\end{cor} 
{\bf Proof\,:} Due to (\ref{madu}) we have 
$G''(\eps)=\sigma (\sigma+1)\,\eps^{\sigma-1}\,\Phi_{2, \sigma-1}(\theta)$, where 
\[ \theta=\sqrt{\frac{\alpha^2}{\eps^2}-1}\in [\theta_q, 1],
   \quad\theta_q=\sqrt{\frac{1}{q^2}-1}\in ]0, 1]. \] 
Then 
\begin{eqnarray*} 
  |\theta-\tilde{\theta}|
  & = & \Big|\sqrt{\frac{\alpha^2}{\eps^2}-1}-\sqrt{\frac{\alpha^2}{\tilde{\eps}^2}-1}\Big|
  =\frac{\alpha^2}{\theta+\tilde{\theta}}\,\frac{|\eps^2-\tilde{\eps}^2|}{\eps^2\tilde{\eps}^2}
  \\ & \le & \frac{4\alpha^2}{2\theta_q}\,\frac{2q\alpha\,|\eps-\tilde{\eps}|}{\alpha^4}
  \le\frac{4}{\theta_q}\,\alpha^{-1}\,|\eps-\tilde{\eps}|. 
\end{eqnarray*} 
Let $C_\ast>0$ denote the constant from Corollary \ref{PhiLip}. Since $\Phi_{2, \sigma-1}(0)=0$ 
by Lemma \ref{Philem}, Corollary \ref{PhiLip} in particular implies that 
$|\Phi_{2, \sigma-1}(\theta)|\le C_\ast\theta\le C_\ast$ for $\theta\in [0, 1]$. 
Hence, using this Corollary \ref{PhiLip}, 
\begin{eqnarray*} 
   |G''(\eps)-G''(\tilde{\eps})|
   & \le & \sigma (\sigma+1)\,|\Phi_{2, \sigma-1}(\theta)|
   \,|\eps^{\sigma-1}-\tilde{\eps}^{\sigma-1}|
   \\ & & +\,\sigma (\sigma+1)\,\tilde{\eps}^{\sigma-1}\,|\Phi_{2, \sigma-1}(\theta)
   -\Phi_{2, \sigma-1}(\tilde{\theta})|
   \\ & \le & C_\ast\,\sigma (\sigma+1)|\sigma-1|\,\Big(\frac{\sqrt{2}}{\alpha}\Big)^{2-\sigma}
   \,|\eps-\tilde{\eps}|
   +C_\ast\,\sigma (\sigma+1)\,\tilde{\eps}^{\sigma-1}
   \,|\theta-\tilde{\theta}|
   \\ & \le & C_\ast\,\sigma (\sigma+1)\,\Big[
   |\sigma-1|\,\Big(\frac{\sqrt{2}}{\alpha}\Big)^{2-\sigma}
   +\frac{4}{\theta_q}\,(\max_{u\in [\frac{1}{\sqrt{2}}\alpha, \alpha]} u^{\sigma-1})
   \,\alpha^{-1}\Big]\,|\eps-\tilde{\eps}|, 
\end{eqnarray*} 
which shows that $C_q$ can be chosen appropriately. 
{\hfill$\Box$}\bigskip

\begin{remark}\label{roi}
{\rm Note that the constants that realize the equivalences (\ref{abm}), (\ref{adv}), (\ref{tawe}), 
in the sense of an upper and a lower bound, are in fact independent of $\alpha$, 
since they only rely on the asymptotics of the $\Phi_{a, b}$ from Lemma \ref{Philem}. 
For instance, if we have 
\begin{eqnarray*}
   & & c_1\theta^{5+2\sigma}\le\Phi_{0, \sigma+1}(\theta)\le c_2\theta^{5+2\sigma},\quad\theta\in [0, 1], 
   \\[1ex] & & c_3\theta^{4+\sigma}\le\Phi_{0, \sigma+1}(\theta)\le c_4\theta^{4+\sigma},
   \quad\theta\in [1, \infty[, 
\end{eqnarray*}  
for constants $c_2>c_1>0$ and $c_4>c_3>0$, then, following the proof of Lemma \ref{Geasy}, 
it is not difficult to see that 
\begin{eqnarray*} 
   & & c_1\alpha^{-3/2}\,(\alpha-\eps)^{5/2+\sigma}\le G(\eps)\le 2^{3/4}(1+\sqrt{2})^{5/2+\sigma} c_2
   \,\alpha^{-3/2}\,(\alpha-\eps)^{5/2+\sigma},\quad\eps\in\Big[\frac{1}{\sqrt{2}}\,\alpha, \alpha\Big], 
   \\ & & \Big(1-\frac{1}{\sqrt{2}}\Big)^{2+\sigma/2} 
   c_3\,\eps^{-3}\,\alpha^{4+\sigma}\le G(\eps)
   \le\Big(1+\frac{1}{\sqrt{2}}\Big)^{2+\sigma/2}\,c_4\,\eps^{-3}\,\alpha^{4+\sigma},
   \quad\eps\in\Big]0, \frac{1}{\sqrt{2}}\,\alpha\Big], 
\end{eqnarray*}    
is obtained 
{\hfill$\diamondsuit$}
}
\end{remark}
\medskip

Let ${\cal G}(s)=\hat{G}(-s)$. 

\begin{lemma}[Properties of ${\cal G}$]\label{calGprops} We have 
\[ {\cal G}(0)=0,
   \quad {\cal G}'(0)=-\alpha,
   \quad\lim_{s\to\infty} {\cal G}'(s)=0,
   \quad\lim_{s\to\infty} {\cal G}(s)=-\infty, \]
and ${\cal G}(s)<0$ for $s\in ]0, \infty[$ as well as ${\cal G}'(s)\in ]-\alpha, 0[$ 
for $s\in ]0, \infty[$. Furthermore, ${\cal G}''(s)>0$ for $s\in ]0, \infty[$ and 
\[ {\cal G}''(s)\sim\alpha^{\frac{3}{3+2\sigma}}
   \,s^{-\frac{1+2\sigma}{3+2\sigma}},\quad s\to 0^+ . \] 
More precisely, let $c_1^\ast, c_2^\ast, c_3^\ast, c_4^\ast, c_5^\ast>0$ 
be constants (independent of $\alpha$) such that 
\begin{equation}\label{drogat} 
   -c_2^\ast\alpha^{-3/2}\,(\alpha-\eps)^{3/2+\sigma}
   \le G'(\eps)\le -c_1^\ast\alpha^{-3/2}\,(\alpha-\eps)^{3/2+\sigma},
   \quad\eps\in\Big[\frac{1}{\sqrt{2}}\,\alpha, \alpha\Big],
\end{equation} 
and 
\[ G'(\eps)\le -c_3^\ast\eps^{-4}\alpha^{4+\sigma},\quad
   \eps\in\Big]0, \frac{1}{\sqrt{2}}\,\alpha\Big], \]  
and moreover 
\begin{equation}\label{kungat} 
   c_4^\ast\alpha^{-3/2}\,(\alpha-\eps)^{1/2+\sigma}
   \le G''(\eps)\le c_5^\ast\alpha^{-3/2}\,(\alpha-\eps)^{1/2+\sigma},
   \quad\eps\in\Big[\frac{1}{\sqrt{2}}\,\alpha, \alpha\Big],
\end{equation} 
are verified; recall (\ref{adv}) and (\ref{tawe}). 
Put $s_0=2c_3^\ast\alpha^\sigma$. Then 
\begin{equation}\label{phca2} 
   (c_5^\ast)^{-1}(c_1^\ast)^{\frac{1+2\sigma}{3+2\sigma}}
   \,\alpha^{\frac{3}{3+2\sigma}}\,s^{-\frac{1+2\sigma}{3+2\sigma}}
   \le {\cal G}''(s)\le (c_4^\ast)^{-1}(c_2^\ast)^{\frac{1+2\sigma}{3+2\sigma}}
   \,\alpha^{\frac{3}{3+2\sigma}}\,s^{-\frac{1+2\sigma}{3+2\sigma}},
\end{equation} 
for all $s\in ]0, s_0]$. 
\end{lemma} 
{\bf Proof\,:} See Section \ref{Legrem} for the notation. 
Since $\inf_{\eps\in\R} G(\eps)=0$, we get ${\cal G}(0)=\hat{G}(0)=-\inf_{\eps\in\R} G(\eps)=0$. 
For the derivative, if $u<0$, then $\hat{G}(u)$ is attained at a unique $\eps(u)\in ]0, \alpha[$. 
As $\eps(u)=(G')^{-1}(u)$ is the inverse function of $G'$, we obtain 
$\lim_{u\to -\infty}\eps(u)=0$ and $\lim_{u\to 0^-}\eps(u)=\alpha$ 
from Lemma \ref{Gesasy}. For $u\to 0^-$ we can hence use (\ref{abm}) and (\ref{adv}) to deduce 
\[ \frac{G(\eps(u))}{u}=\frac{G(\eps(u))}{G'(\eps(u))}
   \sim\frac{\alpha^{-3/2}\,(\alpha-\eps(u))^{5/2+\sigma}}{-\alpha^{-3/2}\,(\alpha-\eps(u))^{3/2+\sigma}}
   =-(\alpha-\eps(u))\to 0,  \]  
so that also $\lim_{u\to 0^-}\frac{G(\eps(u))}{u}=0$. It follows that 
\[ \lim_{u\to 0^-}\frac{\hat{G}(0)-\hat{G}(u)}{-u}
   =\lim_{u\to 0^-}\frac{\hat{G}(u)}{u}=\lim_{u\to 0^-}\Big(\eps(u)-\frac{G(\eps(u))}{u}\Big)=\alpha, \]  
which means that ${\cal G}'(0)=-\hat{G}'(0)=-\alpha$ does exist. Next, due to $\lim_{u\to -\infty}\hat{G}'(u)
=\lim_{u\to -\infty}\eps(u)=0$, we also have $\lim_{s\to\infty} {\cal G}'(s)=0$.  
If $s\in ]0, \infty[$, then $u=-s\in ]-\infty, 0[$ and therefore $\eps(u)\in ]0, \alpha[$. 
Since ${\cal G}''(s)=\frac{1}{G''(\eps(u))}$ by (\ref{nwt}), Lemma \ref{Gessasy} implies that 
${\cal G}''(s)>0$. In particular, ${\cal G}'$ is increasing from $-\alpha$ to $0$, 
which shows that ${\cal G}'(s)\in ]-\alpha, 0[$ for $s\in ]0, \infty[$. As a consequence of ${\cal G}(0)=0$, 
this in turn yields ${\cal G}(s)<0$ for $s\in ]0, \infty[$. To verify 
$\lim_{s\to\infty}{\cal G}(s)=\lim_{u\to -\infty}\hat{G}(u)=-\infty$, we use (\ref{adv}) to deduce 
$G'(\eps)\sim -\eps^{-4}\alpha^{4+\sigma}$ as $\eps\to 0^+$. Since $\eps(u)\to 0^+$ as $u\to -\infty$, 
this gives $u=G'(\eps(u))\sim -\eps(u)^{-4}\alpha^{4+\sigma}$ as $u\to -\infty$, and hence 
\[ \eps(u)u\sim\Big(\frac{\alpha^{4+\sigma}}{-u}\Big)^{1/4}u=-\alpha^{1+\sigma/4}(-u)^{3/4} \] 
as $u\to -\infty$. Also $\eps(u)\to 0^+$ in conjunction with (\ref{abm}) shows that $G(\eps(u))\to\infty$ 
as $u\to -\infty$. Thus $\hat{G}(u)=\eps(u)u-G(\eps(u))\to -\infty$ as $u\to -\infty$. 
To establish (\ref{phca2}), fix $s\in ]0, s_0]$ and let again $u=-s$. 
Suppose that $\eps(u)\le\frac{1}{\sqrt{2}}\,\alpha$. 
Then $u=G'(\eps(u))\le -c_3^\ast\eps(u)^{-4}\alpha^{4+\sigma}$, which implies that 
\[ \frac{1}{4}\,\alpha^4\ge\eps(u)^4\ge c_3^\ast\alpha^{4+\sigma}\,\frac{1}{(-u)}
   \ge c_3^\ast\alpha^{4+\sigma}\,\frac{1}{s_0}=\frac{1}{2}\,\alpha^4, \] 
which is a contradiction. Therefore we must have $\eps(u)\in [\frac{1}{\sqrt{2}}\,\alpha, \alpha]$, 
and (\ref{drogat}) applies. It follows that 
\[ -c_2^\ast\alpha^{-3/2}\,(\alpha-\eps(u))^{3/2+\sigma}
   \le u\le -c_1^\ast\alpha^{-3/2}\,(\alpha-\eps(u))^{3/2+\sigma}, \] 
and consequently 
\begin{equation}\label{barb} 
   (c_2^\ast)^{-\frac{2}{3+2\sigma}}\,\alpha^{\frac{3}{3+2\sigma}}\,(-u)^{\frac{2}{3+2\sigma}}
   \le\alpha-\eps(u)
   \le (c_1^\ast)^{-\frac{2}{3+2\sigma}}\,\alpha^{\frac{3}{3+2\sigma}}\,(-u)^{\frac{2}{3+2\sigma}}.
\end{equation} 
From (\ref{kungat}) and ${\cal G}''(s)=\frac{1}{G''(\eps(u))}$ we deduce that 
\[ (c_5^\ast)^{-1}\alpha^{3/2}\,(\alpha-\eps(u))^{-(1/2+\sigma)}
   \le {\cal G}''(s)\le (c_4^\ast)^{-1}\alpha^{3/2}\,(\alpha-\eps(u))^{-(1/2+\sigma)}, \] 
and hence we can use (\ref{barb}) to get 
\[ (c_5^\ast)^{-1}(c_1^\ast)^{\frac{1+2\sigma}{3+2\sigma}}
   \,\alpha^{3/2}\,\alpha^{-\frac{3(1+2\sigma)}{2(3+2\sigma)}}\,{(-u)}^{-\frac{1+2\sigma}{3+2\sigma}}
   \le {\cal G}''(s)\le (c_4^\ast)^{-1}(c_2^\ast)^{\frac{1+2\sigma}{3+2\sigma}}
   \,\alpha^{3/2}\,\alpha^{-\frac{3(1+2\sigma)}{2(3+2\sigma)}}\,{(-u)}^{-\frac{1+2\sigma}{3+2\sigma}}, \] 
as asserted. 
{\hfill$\Box$}\bigskip 

\subsection{Legendre transforms} 
\label{Legrem}

The Legendre transform of a general strictly convex real-valued function $G$ is 
\[ \hat{G}(u)=\sup_{\eps\in\R}\,(\eps u-G(\eps))
   =-\inf_{\eps\in\R}\,(G(\eps)-\eps u); \] 
see \cite[Section 8.1]{FasanoMarmi}. 
For every fixed $u$ let $\eps(u)\in\R$ be such that 
$\hat{G}(u)=\eps(u)u-G(\eps(u))$ and 
\[ 0=\frac{d}{d\eps}\,[\eps u-G(\eps)]\Big|_{\eps=\eps(u)}
   =u-G'(\eps(u)). \]  
From the strict convexity of $G$ it follows that $G'$ is invertible. 
Therefore $\eps(u)=(G')^{-1}(u)$, and then 
\[ \hat{G}'(u)=\eps(u)+\eps'(u)u-G'(\eps(u))\eps'(u)
   =\eps(u) \] 
shows that 
\begin{equation}\label{derLeg}
   \hat{G}'(u)=\eps(u)=(G')^{-1}(u)
\end{equation} 
is the derivative of the Legendre transform. 
\smallskip

The relation $\widehat{\widehat{G}}=G$ is also useful.  
To calculate the Legendre transform of $\hat{G}$, one first has to locate $u(\eps)$ 
such that $0=\eps-\hat{G}'(u(\eps))$, and then $\widehat{\widehat{G}}(\eps)
=u(\eps)\eps-\hat{G}(u(\eps))$. Thus $u(\eps)=G'(\eps)$ is found, 
which leads to 
\[ \widehat{\widehat{G}}(\eps)=G'(\eps)\eps-\hat{G}(G'(\eps))
   =G'(\eps)\eps-\eps(G'(\eps))G'(\eps)+G(\eps(G'(\eps)))=G(\eps) \] 
as claimed, using that $\eps(G'(\eps))=(G')^{-1}(G'(\eps))=\eps$.  
\smallskip 

Next let ${\cal G}(s)=\hat{G}(-s)$. Then 
\begin{equation}\label{dozu} 
   {\cal G}(s)-s\,{\cal G}'(s)=-G(-{\cal G}'(s)).
\end{equation} 
To establish (\ref{dozu}), we write $u=-s$ and get from ${\cal G}'(s)=-\hat{G}'(-s)$ that 
${\cal G}(s)-s\,{\cal G}'(s)=\hat{G}(u)-u\,\hat{G}'(u)=\hat{G}(u)-\eps(u)u=-G(\eps(u))$, 
and thus the claim follows from $\eps(u)=\hat{G}'(u)=\hat{G}'(-s)=-{\cal G}'(s)$.   
\smallskip 

Another noteworthy relation is 
\begin{equation}\label{nwt} 
   {\cal G}''(s)=\frac{1}{G''(\eps(-s))}.
\end{equation} 
In fact, ${\cal G}''(s)=\hat{G}''(-s)=\hat{G}''(u)$, and (\ref{derLeg}) yields 
$G'(\hat{G}'(u))=u$, which upon differentiation shows that 
$1=G''(\hat{G}'(u))\hat{G}''(u)=G''(\eps(u))\hat{G}''(u)$. 


\end{document}